\documentclass{ptephy}

\makeatletter
\@addtoreset{equation}{section}

\makeatother

\newcommand{\refer}[1]{(\ref{#1})}
\newcommand{\oper}[1]{\mathcal{#1}}
\newcommand{\Tr}[1]{\mathrm{Tr}\left(#1\right)}
\newcommand{\abs}[1]{\left|#1\right|}
\newcommand{\comm}[2]{\left[ #1, #2 \right]}

\newcommand{\diff}{
{d}}

\newcommand{\lineint}{\int\limits_{-\infty}^{\infty}\diff}

\renewcommand{\Re}{\mathrm{Re}}
\renewcommand{\Im}{\mathrm{Im}}

\def\beq{\begin{eqnarray}}
\def\eeq{\end{eqnarray}}
\def\Tr{{\rm Tr}}

\begin{document}

\title{
Stabilizing matter and gauge fields localized on walls 
}

\author{
\name{\fname{Masato } \surname{Arai}}{1,2,\ast},
\name{\fname{Filip} \surname{Blaschke}}{2,}$^{**}$, 
\name{\fname{Minoru} \surname{Eto}}{3,}$^{***}$, and
\name{\fname{Norisuke} \surname{Sakai}}{4,}$^{****}$}

\address{\affil{1}{Fukushima National College of Technology, 
Iwaki, Fukushima 970-8034, Japan}
\affil{2}{Institute of Experimental and Applied Physics, 
Czech Technical University in Prague, Horsk\'a 22, 128 00 
Prague 2, Czech Republic, and 
Institute of Physics, Silesian University in Opava, 
Bezru\v{c}ovo n\'am. 1150/13, 746~01 Opava, Czech Republic}
\affil{3}{Department of Physics, Yamagata University, Yamagata 
990-8560, Japan}
\affil{4}{Department of Mathematics, Tokyo Woman's 
Christian University, Tokyo 167-8585, Japan}
\email{masato.arai@fukushima-nct.ac.jp, 
$^{**}$Email: filip.blaschke@fpf.slu.cz, \\
$^{***}$Email: meto@sci.kj.yamagata-u.ac.jp 
$^{****}$Email: norisuke.sakai@gmail.com
}
}

\begin{abstract}%
Both non-Abelian gauge fields and minimally interacting 
massless matter fields are localized on a domain wall in the 
five-dimensional spacetime. 
Field-dependent gauge coupling naturally gives a 
position-dependent coupling to localize non-Abelian gauge 
fields on the domain wall. 
An economical field content allows us to 
eliminate a moduli for a instability, and 
to demonstrate the positivity of the position-dependent 
coupling in the entire moduli space. 
Effective Lagrangian similar to the chiral Lagrangian is 
found with a new feature of different coupling strengths 
for adjoint and singlet matter that depend on the width 
of the domain wall. 
\end{abstract}

\maketitle

\section{Introduction}\label{sc:introduction}

Localization of massless gauge fields on a domain wall 
with $3+1$ dimensional world volume has been a 
long-standing problem to achieve the dynamical 
compactification in the brane-world scenario \cite{DuRu}. 
If the gauge symmetry is unbroken only inside the domain 
wall and broken outside, the gauge field inevitably 
acquires a mass 
proportional to  
the inverse of the width of the wall
\cite{DaSh,Antoniadis:1998ig,Maru:2003mx}. 
The localization of gauge fields can be achieved if the 
gauge theory is in the confining phase outside of the 
domain wall \cite{DaSh,Antoniadis:1998ig}. 
This requirement can be translated into a position-dependent 
gauge coupling \cite{Kogut:1974sn, Fukuda:1977wj,KaKu}. 

The supersymmetric gauge theories in $4+1$ spacetime 
dimensions allows a cubic coupling \cite{Seiberg:1996bd} 
between adjoint scalar 
fields $\Sigma^\alpha$ and gauge field strengths 
$F_{MN}^\beta$ 
\begin{equation}
{\mathcal L} \sim C_{\alpha\beta\gamma}\Sigma^\alpha 
F^\beta_{MN}F^{\gamma MN},
\label{eq:general_cubic_coupling}
\end{equation}
with 
coupling constants
$C_{\alpha\beta\gamma}$. 
In Ref.\cite{Ohta:2010fu}, a domain wall solution 
is chosen 
so that the scalar field 
$\Sigma$  
is positive
inside and  
vanishes
asymptotically outside of the domain wall.
In this way, non-Abelian gauge fields 
have
been localized on the domain wall. 
More recently a method to localize non-Abelian 
gauge fields together with minimally interacting matter 
fields on domain walls has been introduced and a particular 
model has been presented in five-dimensional spacetime 
\cite{Arai:2012cx, Arai:2012dm}, by gauging the unbroken 
global symmetry 
associated to the degenerate scalar fields to form the 
domain wall \cite{Shifman:2003uh, Eto:2005cc, Eto:2008dm}. 
This mechanism provides a step towards a 
realistic model of branes as soliton solutions of 
higher dimensional field theories, and realizes 
one of the most important characteristics of D-branes: 
massless $U(N)$ gauge fields emerge when $N$ walls 
are coincident, and become massive as walls separate.

\begin{figure}[ht]
\begin{center}
\begin{minipage}[b]{0.49\linewidth}
\centering
\includegraphics[width=\textwidth]{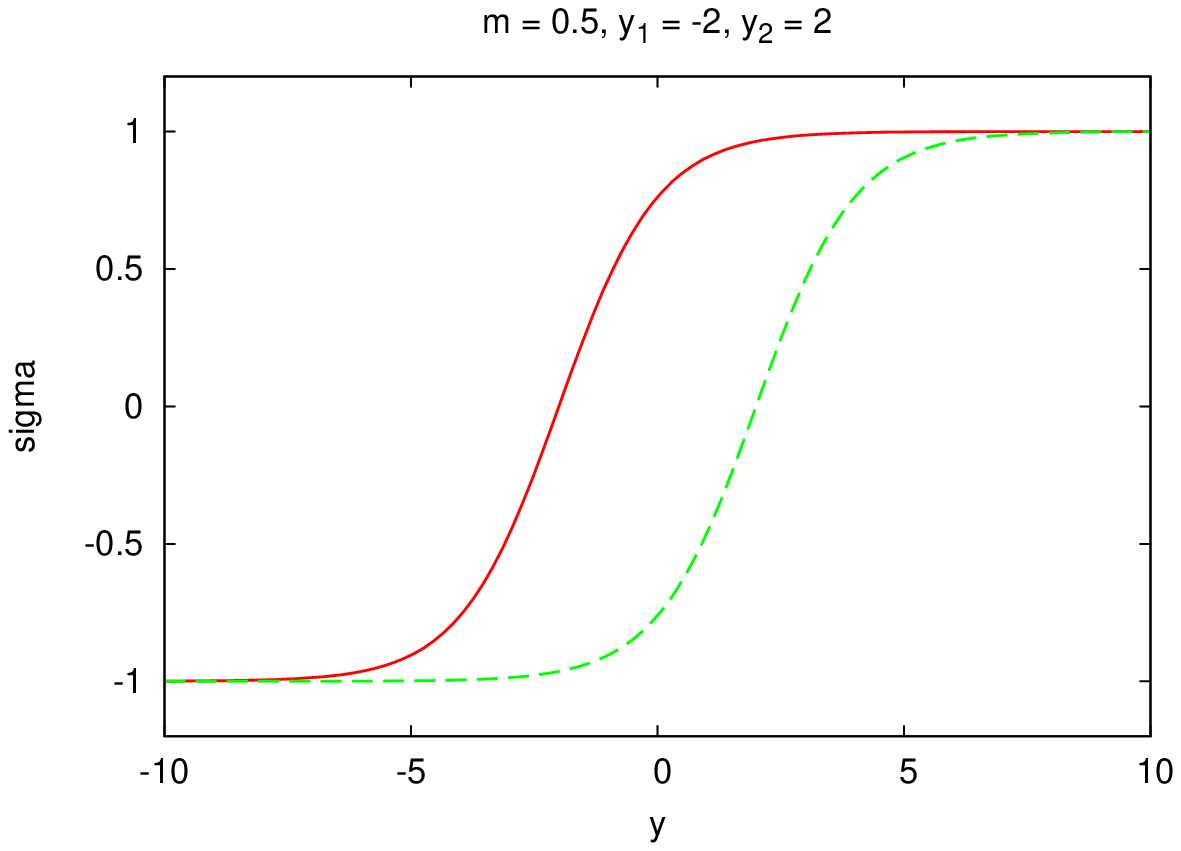}
\end{minipage}
\begin{minipage}[b]{0.49\linewidth}
\centering
\includegraphics[width=\textwidth]{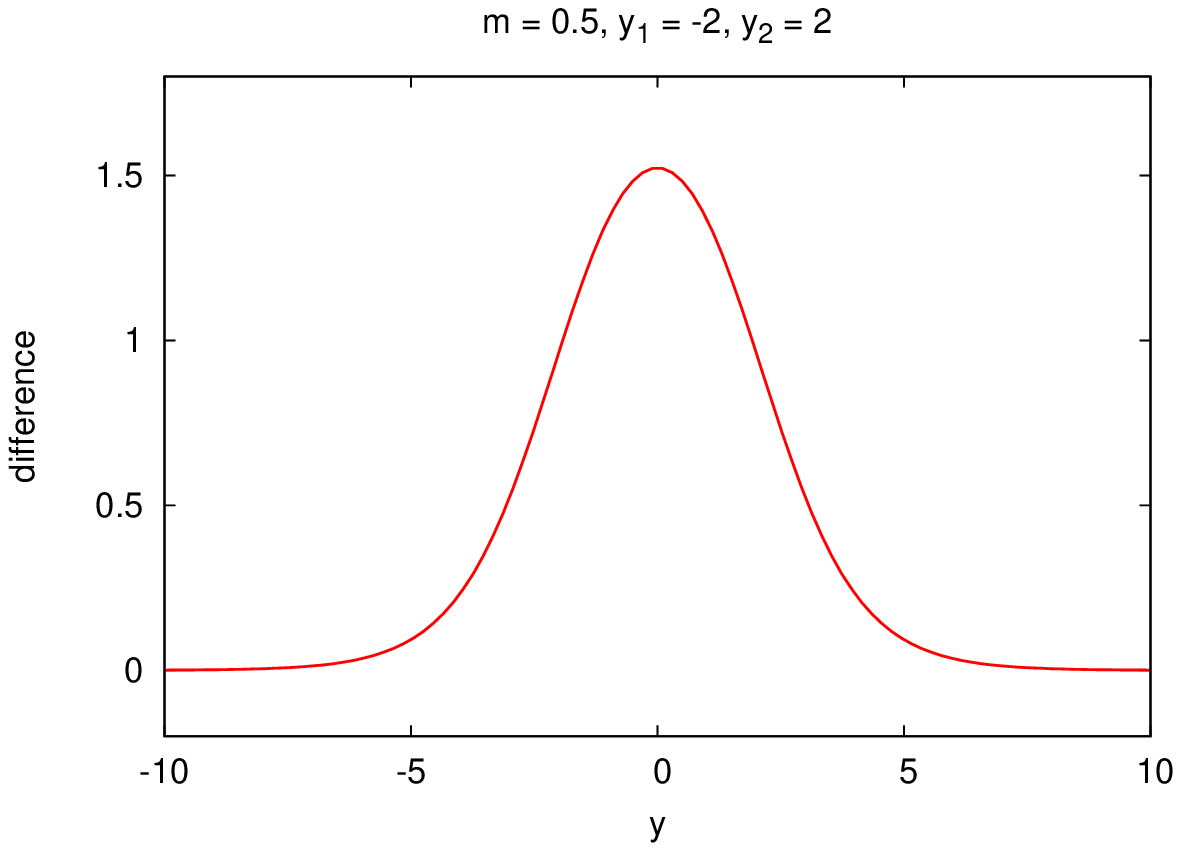}
\end{minipage}
\caption{Position-dependent gauge coupling 
(right panel) is given as a difference $\sigma_1-\sigma_2$
of two separated kinks (left panel).}
\label{fig:01}
\end{center}
\end{figure}

In Ref.~\cite{Ohta:2010fu}, two real scalar fields $\sigma_1$ 
and $\sigma_2$ were introduced to form the usual kink profile 
in the extra-dimensional coordinate $y$ as shown in 
the left panel of Fig. \ref{fig:01}, and the cubic 
coupling in Eq.(\ref{eq:general_cubic_coupling}) 
is chosen as 
\begin{eqnarray}
{\mathcal L}_{\rm cubic} \sim - (\sigma_1 - \sigma_2) 
\Tr[\tilde G_{MN}\tilde G^{MN}],
\label{eq:CS}
\end{eqnarray}
where $\tilde G_{MN}$ is the non-Abelian gauge field 
strength to be localized on the domain wall. 
In this way, one achieved the desired profile of the 
position-dependent coupling which is positive and vanishes 
asymptotically outside of the domain wall as shown in the 
right panel of Fig.\ref{fig:01}. 
More recently, matter fields have also been localized on 
the domain wall interacting minimally with the localized 
gauge fields \cite{Arai:2012cx}. 
Although the model realized the localization mechanism 
in a simple setup, the width of the position-dependent 
coupling is a modulus and can become negative implying 
an instability of the gauge kinetic term. 

Moduli of domain walls arise as remaining degrees 
of freedom of scalar fields constrained by gauge symmetry 
to form the domain wall. 
This observation prompts us to construct a model with smaller 
number of scalar fields with a different charge assignments 
in order to eliminate the unwanted modulus in the 
position-dependent coupling \cite{Ohta:2010fu}.

The purpose of this paper is to present models in $4+1$ 
dimensional spacetime with domain wall solutions which 
allow stable localized gauge fields interacting minimally 
with localized matter fields. 
For any values of the moduli parameters, we show that 
the position-dependent gauge coupling is positive 
everywhere in the extra-dimensional coordinate $y$. 
In particular, its width is no longer a modulus and is 
fixed by the parameters of the theory. 
We thus find that the gauge kinetic term is stable 
for all values of moduli, namely for all possible 
domain wall configurations. 
As localized matter fields, we obtain scalar fields 
in singlet and adjoint representations of the gauge group. 
They are associated with the broken part of the global 
symmetry, and interact minimally with the localized gauge 
fields. 
The effective Lagrangian on the domain wall is also worked 
out, using the method in Ref.\cite{Eto:2006uw}. 
It resembles the chiral Lagrangian associated with the chiral 
symmetry breaking of QCD. 
A new feature of the effective Lagrangian is that the coupling 
strength of the adjoint matter fields is larger than that of 
the singlet matter fields. 
We examine generality of models with 
a
stable kinetic term for gauge fields to identify the class 
of models with desirable properties. 
We also discuss ways to build more realistic models 
for a phenomenology of the brane-world scenario.

The organization of the paper is as follows.
Section 2 is devoted to construct a model in $4+1$ 
spacetime dimensions with domain wall solutions which 
localizes matter and gauge fields. 
The position-dependent gauge coupling is shown to be 
positive. 
In Section 3, the low-energy effective Lagrangian is 
obtained. 
In Section 4, we examine the generality of models with 
the stable gauge kinetic term. 
Conclusion and discussion are in Section 5. 
Some details to derive the low-energy 
effective Lagrangian is given in Appendix A.
In Appendix B, we summarize geometrical features of the three-flavor model.

\section{
A model with localized matter and gauge fields 
}\label{sc:threeflavor}

In this section we first present a model allowing the domain 
wall solution with unbroken non-Abelian global symmetry 
\cite{Eto:2008dm}. 
As the second step, we introduce non-Abelian gauge fields 
for the unbroken symmetry. 
As the third step, we consider the cubic coupling 
\cite{Ohta:2010fu} in Eq.(\ref{eq:general_cubic_coupling}) 
to localize non-Abelian gauge fields : 
we use expectation values of a singlet scalar  field 
to give the position-dependent gauge coupling, whose 
positivity is demonstrated for any values of moduli 
parameters. 
A
broken part of the global symmetry provides matter fields 
in singlet and adjoint representations of the localized 
gauge fields \cite{Arai:2012cx}.

\subsection{Lagrangian with global symmetry 
and domain wall solutions}

Let us consider a five-dimensional 
$SU(N)_c\times U(1)_1\times U(1)_2$ gauge theory 
and $N$ scalar fields $H_1$ ($H_2$) in the fundamental 
representation with the degenerate mass $m$ ($-m$), 
together with a singlet scalar field $H_3$, whose charge 
assignments are summarized in Tab.~\ref{table:01}. 
Therefore we obtain global symmetry 
$SU(N)_L\times SU(N)_R\times U(1)_A$. 
In addition, we introduce adjoint and singlet scalars 
$\Sigma$ and $\sigma$ associated with the gauge group 
$SU(N)_c\times U(1)_1$ and $U(1)_2$, respectively. 
\begin{table}
\begin{center}
\begin{tabular}{c|ccc|ccc|c}
\hline
  & $SU(N)_{c}$ & $U(1)_1$ & $U(1)_2$ & $SU(N)_{L}$ 
& $SU(N)_{R}$ & $U(1)_{A}$ & mass\\ \hline
$H_{1}$ & $\square$ & 1 & 0 & $\square$ & {\bf 1} 
& 1 & $m{\bf 1}_{N}$\\ 
$H_{2}$ & $\square$ &  1 & $-1$ & {\bf 1} 
& $\square$ & $-1$ & 0 \\ 
$H_{3}$ & {\bf 1} & 0 & 1 & {\bf 1} 
& {\bf 1} & 0  & 0\\ 
$\Sigma$ & ${\rm adj}\oplus{\bf 1}$ & 0 & 0 & {\bf 1} 
& {\bf 1} & 0 & 0\\
$\sigma$ & ${\bf 1}$ & 0 & 0 & {\bf 1} & {\bf 1} & 0 & 0 \\
\hline
\end{tabular}
\end{center}
\caption{Quantum numbers of fields of the model for 
the domain wall. 
}
\label{table:01}
\end{table}
We assume the following Lagrangian 
with the signature $(+,-,-,-,-)$ 
\begin{align}
\oper{L}   
& =-\frac{1}{2g^2}\Tr\Bigl(G_{MN}G^{MN}\Bigr) 
-\frac{1}{4e^2}F_{MN}F^{MN} 
+\frac{1}{g^2}\Tr\Bigl(D_M\Sigma\Bigr)^2
 +\frac{1}{2e^2}(\partial_M\sigma)^2\nonumber \\ 
\label{eq:lag1}
 & +\Tr\abs{D_MH_1}^2+\Tr\abs{(D_M-i A_M)H_2}^2
+\abs{(\partial_M+i A_M)H_3}^2 -V\,, \\
V & = \Tr\abs{(\Sigma-m\mathbf{1}_N)H_1}^2  
+ \Tr\abs{(\Sigma-\sigma\mathbf{1}_N)H_2}^2
+\abs{\sigma H_3}^2 
\nonumber \\ 
\label{eq:pot1}
& +\frac{1}{4}g^2\Tr\Bigl(c_1\mathbf{1}_N-H_1H_1^{\dagger}
-H_2H_2^{\dagger}\Bigr)^2 + 
\frac{1}{2}e^2\Bigl(c_2+\Tr(H_2H_2^{\dagger})
-\abs{H_3}^2\Bigr)^2\,.
\end{align}
The $U(N)_c = SU(N)_c\times U(1)_1$ gauge coupling and 
gauge fields are denoted by $g$ and an $N\times N$ matrix 
$W_M$ with $M=0,1,2,3,4$.  
The $U(1)_2$ gauge coupling and gauge field are denoted 
by $e$ and $A_M$. 
Covariant derivatives and field strengths are defined by 
\begin{equation}
\label{eq:cov1} 
D_M H_{1,2}  = \partial_M H_{1,2} +i W_{M} H_{1,2}, 
\quad 
D_M \Sigma  = \partial_M \Sigma +i \comm{W_M}{\Sigma}\,, 
\end{equation}
and $G_{MN} = \partial_MW_N-\partial_NW_M+i \comm{W_M}{W_N}$, 
$F_{MN} = \partial_M A_N-\partial_NA_M$. 

The global symmetry $U_L \in SU(N)_L, U_R \in SU(N)_R$, 
$e^{i \alpha} \in U(1)_A$, 
and 
the
local gauge symmetry $U_c \in U(N)_c$,
$e^{i \beta} \in U(1)_2$
act on the fields as 
\begin{gather}
\label{eq:symm1} H_1 \to e^{i \alpha} U_c H_1 U_L \,, \hspace{0.5cm} H_2 \to e^{-i (\alpha+\beta)}U_c H_2 U_R \,, \hspace{0.5cm}
H_3 \to e^{i \beta} H_3\,, \\
\label{eq:symm2} \Sigma \to U_c\Sigma U_c^{\dagger}\,, \hspace{1cm} \sigma \to \sigma\,.
\end{gather}
Let us note that the Lagrangian \refer{eq:pot1} can be 
embedded into the five-dimensional ${\mathcal N}=1$ 
supersymmetric gauge theory (with $8$ supercharges). 
This fact allows us to obtain the so-called 
Bogomol'nyi-Prasad-Sommerfield (BPS) domain wall solution 
as we will show in the next subsection. 
We stress, however, that this particular choice is merely 
to simplify calculations and is not required for our results 
to hold.

Without loss of generality, we can assume the mass 
parameter $m$ to be positive. 
We also assume $c_1 > 0$ and $c_2 > 0$. 
Then there exist $N+1$ discrete vacua with 
 $r =0,1,\,\ldots\,, N$, where scalar fields 
 develop vacuum expectation value (VEV) 
\begin{gather}\label{eq:vev}
H_1 = \sqrt{c_1}\begin{pmatrix}
\mathbf{1}_{N-r} & \\
& \mathbf{0}_{r}\end{pmatrix}\,, \hspace{0.5cm}
H_2 = \sqrt{c_1}\begin{pmatrix}
\mathbf{0}_{N-r} & \\
& \mathbf{1}_{r}\end{pmatrix}\,, \hspace{0.5cm} 
H_3 = \sqrt{c_2+rc_1}\,, \\
\Sigma = m \begin{pmatrix}
\mathbf{1}_{N-r} & \\
 & \mathbf{0}_r\end{pmatrix}\,, \hspace{1cm} \sigma = 0\,.
\end{gather}
Local gauge symmetry is completely broken and only a subgroup 
of global symmetry remains 
in these vacua. 
The breaking patterns in the $r=0$ and $r=N$ vacua are 
\begin{align*}
&U(N)_c\times SU(N)_L\times SU(N)_R\times U(1)_2\times 
U(1)_A \qquad & \\
& \qquad \qquad \xrightarrow[0-\mathrm{th\ vacuum}]{}  
SU(N)_{L+c}\times SU(N)_R\times U(1)_{A+c}\,, \\
& \qquad \qquad \xrightarrow[N-\mathrm{th\ vacuum}]{}  
SU(N)_{R+c}\times SU(N)_L\times U(1)_{A-c}\,.
\end{align*}
Let us consider domain wall solutions connecting $N$-th 
 ($0$-th) vacuum at left (right) infinity 
$y=-\infty~(y=\infty)$. 
Then the (coincident) domain wall solutions preserve 
the diagonal subgroup as the largest global\footnote{
Local gauge symmetry $SU(N)_c$ contains global 
symmetry as a constant gauge transformation, which is 
displayed in the above symmetry breaking pattern of 
$r$-th and $N$-th vacuua. 
However, Nambu-Goldstone modes only come from the 
genuine global symmetry which is not locally gauged. 
Hence we do not count the $SU(N)_c$ transformations 
to preserve the color-flavor-locked vacua (\ref{eq:vev}). 
The discrete symmetry in the denominator 
corresponds to a simultaneous rotation of the 
$U(1)_A$ by an integer multiple of the angle $2\pi/N$ 
and the center $Z_N$ of $SU(N)_{L-R}$. 
Physically it represents indistinguishability of $N$ 
domain walls. 
} symmetry 
$SU(N)_{L+R}$, providing the Nambu-Goldstone modes 
associated to the breaking of the global symmetry 
\begin{equation}
\frac{SU(N)_L\times SU(N)_R\times U(1)_A}  
{SU(N)_{L+R}\times Z_N}.
\label{eq:nambu_goldstone_mode}
\end{equation}
We assume that fields depend only on extra-dimensional 
coordinate $y$ and that all gauge fields except $W_y$ 
and $A_y$ vanish. 

Thanks to the special choice of the potential, 
we can rewrite the energy density as 
\begin{align}
\oper{E} & = \frac{1}{g^2}\Tr\biggl[D_y\Sigma
-\frac{g^2}{2}\Bigl(c_1\mathbf{1}_N-H_1H_1^{\dagger}
-H_2H_2^{\dagger}\Bigr)\biggr]^2 +
\Tr\abs{D_yH_1+(\Sigma-m\mathbf{1}_N)H_1}^2\nonumber \\
& +\frac{1}{2e^2}\biggl(\partial_y\sigma-e^2\Bigl(c_2
+\Tr(H_2H_2^{\dagger}-\abs{H_3}^2)\Bigr)\biggr)^2+
\Tr\abs{D_yH_2+(\Sigma-(\sigma+i A_y)\mathbf{1}_N)H_2}^2 
\nonumber \\
& +\abs{\partial_y H_3 + (\sigma+i A_y)H_3}^2 
+c_2\partial_y \sigma \nonumber \\
& +\partial_y\Tr\biggl[c_1\Sigma -H_1H_1^{\dagger}(\Sigma
-m\mathbf{1}_N)
-H_2H_2^{\dagger}(\Sigma-\sigma\mathbf{1}_N)\biggr]\,.
\end{align}
Thus, we obtain the Bogomol'nyi bound for 
the total energy (per unit world volume) $E$ 
\begin{equation}
E = \lineint y \,\oper{E} \ge T 
= \lineint y\, \left[c_1\partial_y \Tr(\Sigma) 
+c_2\partial_y \sigma \right]= Nmc_1\,,
\end{equation}
where $T$ is the tension of the domain wall. 
This bound is saturated when the following BPS equations 
are satisfied 
\begin{align}
\label{eq:h1} \partial_yH_1+(\Sigma +i W_y - m\mathbf{1}_N)H_1 & = 0\,, \\
\label{eq:h2} \partial_yH_2+\Bigl(\Sigma +i W_y - (\sigma+i A_y)\mathbf{1}_N\Bigr)H_2 & = 0\,, \\
\label{eq:h3} \partial_yH_3+(\sigma+i A_y)H_3 & = 0\,,
\end{align}
\begin{align}
\label{eq:s1} D_y\Sigma & = \frac{1}{2}g^2\Bigl(c_1\mathbf{1}_N-H_1H_1^{\dagger}-H_2H_2^{\dagger}\Bigr)\,, \\
\label{eq:s2} \partial_y\sigma & = e^2\Bigl(c_2+\Tr(H_2H_2^{\dagger})-\abs{H_3}^2\Bigr)\,.
\end{align}

To use the moduli-matrix 
formalism \cite{Isozumi:2004jc,Isozumi:2004va,Eto:2006pg}, 
we introduce 
$S(y)\in GL(N,\mathbb{C})$ and 
$\psi(y) \in \mathbb{C}$ 
\begin{equation}
\label{eq:s01}\Sigma+i W_y 
 = S^{-1}\partial_y S, 
\quad 
\sigma+i A_y 
 = \frac{1}{2}\partial_y \psi\,.
\end{equation}
Then the matter part (\ref{eq:h1})-(\ref{eq:h3}) 
can be solved by  
\begin{align}
\label{eq:h01}H_1 & = e^{my}S^{-1}H_1^0\,, \\
\label{eq:h02}H_2 & = e^{\frac{1}{2}\psi}S^{-1}H_2^0\,, \\
\label{eq:h03}H_3 & = e^{-\frac{1}{2}\psi}H_3^0\,,
\end{align}
with complex constant $N\times N$ moduli matrices 
$H_1^0,H_2^0$ and a moduli constant $H_3^0$, which 
describe moduli of the  
solution.
The rest of the BPS equations (\ref{eq:s1}) and 
(\ref{eq:s2}) turn into the master equations for 
gauge-invariant Hermitian fields $
\Omega \equiv SS^{\dagger}$ and 
$\eta \equiv \Re(\psi)$. 
\begin{align}
\label{eq:master01}\partial_y(\partial_y\Omega\Omega^{-1}) 
& = \frac{1}{2}g^2\Bigl(c_1\mathbf{1}_N
-(e^{2my}H_1^0H_1^{0\, \dagger}
+e^{\eta}H_2^0H_2^{0\, \dagger})\Omega^{-1}\Bigr)\,,\\
\label{eq:master02}\frac{1}{2}\partial_y^2 \eta & 
= e^2\Bigl(c_2+e^{\eta}\Tr(H_2^0H_2^{0\, \dagger}\Omega^{-1})
-e^{-\eta}\abs{H_3^0}^2\Bigr)\,.
\end{align}
Moduli matrices related by the following 
 $V$-transformations give identical physical fields 
\begin{equation}
(S,\psi,H_1^0,H_2^0,H_3^0) \to (VS,\psi+v, VH_1^0, 
VH_2^0e^{-\frac{1}{2}v},H_3^0e^{\frac{1}{2}v})\,,
\end{equation}
where $V\in GL(N,\mathbb{C})$ and $v \in \mathbb{C}$.
The equivalence class quotiented by this $V$-transformation 
defines the moduli space of domain walls. 
We can use this freedom to choose the form of the moduli 
matrices
\begin{equation}
H_1^0 = \sqrt{c_1}\mathbf{1}_N\,, \hspace{1cm} 
H_3^0 = \sqrt{c_2}\,.\label{eq:ms}
\end{equation}
Let us also decompose $H_2^0$ as 
\begin{equation}\label{eq:decomp}
H_2^0 = \sqrt{c_1}e^{\phi}U^{\dagger}\,,
\end{equation}
where $\phi$ is a Hermitian $N\times N$ matrix and $U$ is 
a unitary $N\times N$ matrix. 
With this choice, the master equations (\ref{eq:master01}) 
and (\ref{eq:master02}) become 
\begin{align}
\label{eq:master1}\partial_y(\partial_y\Omega\Omega^{-1}) 
& = \frac{c_1}{2}g^2\Bigl(\mathbf{1}_N-
\Omega_0\Omega^{-1}\Bigr)\,,\\
\label{eq:master2}\frac{1}{2}\partial_y^2 \eta & 
= e^2\Bigl(c_2+c_1e^{\eta}\Tr(e^{2\phi}\Omega^{-1})
-e^{-\eta}c_2\Bigr)\,,
\end{align}
where $\Omega_0 = e^{2my}\mathbf{1}_N+e^{2\phi}e^{\eta}$.

No analytic solution of this system of the differential 
equations is known in general. 
However, one can study essential features of solutions, 
if one takes the strong gauge coupling limit 
$g^2, e^2 \to \infty$. 
Eqs. (\ref{eq:master1}) and (\ref{eq:master2}) reduce to 
a system of algebraic equations in this limit:
\begin{align}
\label{eq:strong1} \Omega & = e^{2my}\mathbf{1}_N
+e^{2\phi}e^{\eta}\,, \\
\label{eq:strong2} c_2 & = e^{-\eta}c_2
-c_1e^{\eta}\Tr(e^{2\phi}\Omega^{-1})\,.
\end{align}

It turns out that the effective theory describing massless 
excitations localized on the background solution of the 
equations of this system precisely coincides (at least in 
the lowest order of approximation) with the one obtained from 
(\ref{eq:master1}) and (\ref{eq:master2}) (see the detailed 
discussion in Ref. \cite{Arai:2012cx}). 
It is therefore sufficient just to study solutions of 
(\ref{eq:strong1}) and (\ref{eq:strong2}). 

The moduli $\phi$ can be diagonalized by a unitary 
matrix $P$ 
\begin{equation}
\phi = m P^{-1}\mathrm{diag}(y_1,\ldots,y_N)P\,.
\end{equation} 
Then \refer{eq:strong2} reduces to a polynomial equation of 
order $N+1$ for $x:= e^{-\eta}$ 
\begin{equation}\label{eq:x}
x = 1 + \frac{c_1}{c_2}\sum_{i=1}^N \frac{1}{1+e_i x},\quad
e_i = e^{2m(y-y_i)}.
\end{equation} 
If this equation is solved, one can supply its solution 
into Eq.\refer{eq:strong1} to obtain $\Omega$.

In the simplest case, where all walls are coincident $\phi = my_0\mathbf{1}_N$, we can solve equation  \refer{eq:x} 
explicitly ($e_0 := e^{2m(y-y_0)}$) to find 
\begin{align}
\label{coin1} e^{-\eta} & = \frac{1}{2e_0}\Bigl(e_0-1+\sqrt{(1-e_0)^2+4(1+Nc_1/c_2)e_0}\Bigr)\,, \\
\label{coin2} \Omega & = (e^{2my} + e^{2my_0}e^{\eta})\mathbf{1}_N\,. 
\end{align}
Physical fields can be expressed in terms of $\Omega$ and 
$\sigma$ as 
\begin{align}
\label{eq:coin1} H_1 & = \sqrt{c_1}\frac{\mathbf{1}_N}{\sqrt{1+e^{-2m(y-y_0)+\eta}}}\,, \\
\label{eq:coin2} H_2 & = \sqrt{c_1}\frac{U^{\dagger}}{\sqrt{1+e^{2m(y-y_0)-\eta}}}\,, \\
\label{eq:coin3} H_3 & = \sqrt{c_2}e^{-\eta/2}\,, \\
\label{eq:coin4} \Sigma & = \frac{1}{2}\partial_y \ln\Omega\,, \\
\label{eq:coin5} \sigma & = \partial_y \eta\,, \\
\label{eq:coin6} W_y & = A_y = 0\,,
\end{align}
where we fixed the gauge such that 
$S = \Omega^{1/2}$ and $\Im (\psi) = 0$.

\begin{figure}
\begin{center}
\begin{minipage}[b]{0.49\linewidth}
\centering
\includegraphics[width=\textwidth]{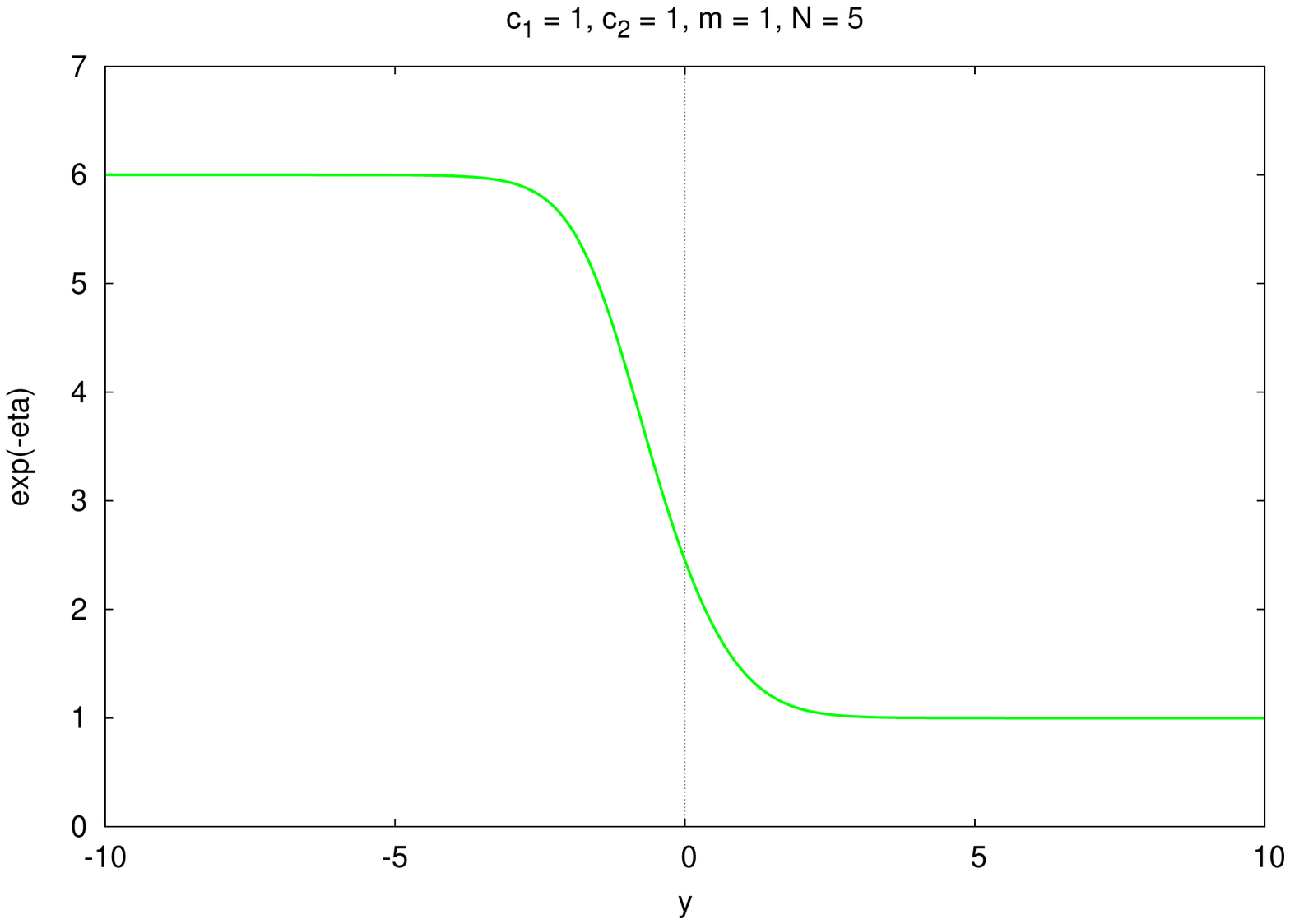}
\end{minipage}
\begin{minipage}[b]{0.49\linewidth}
\centering
\includegraphics[width=\textwidth]{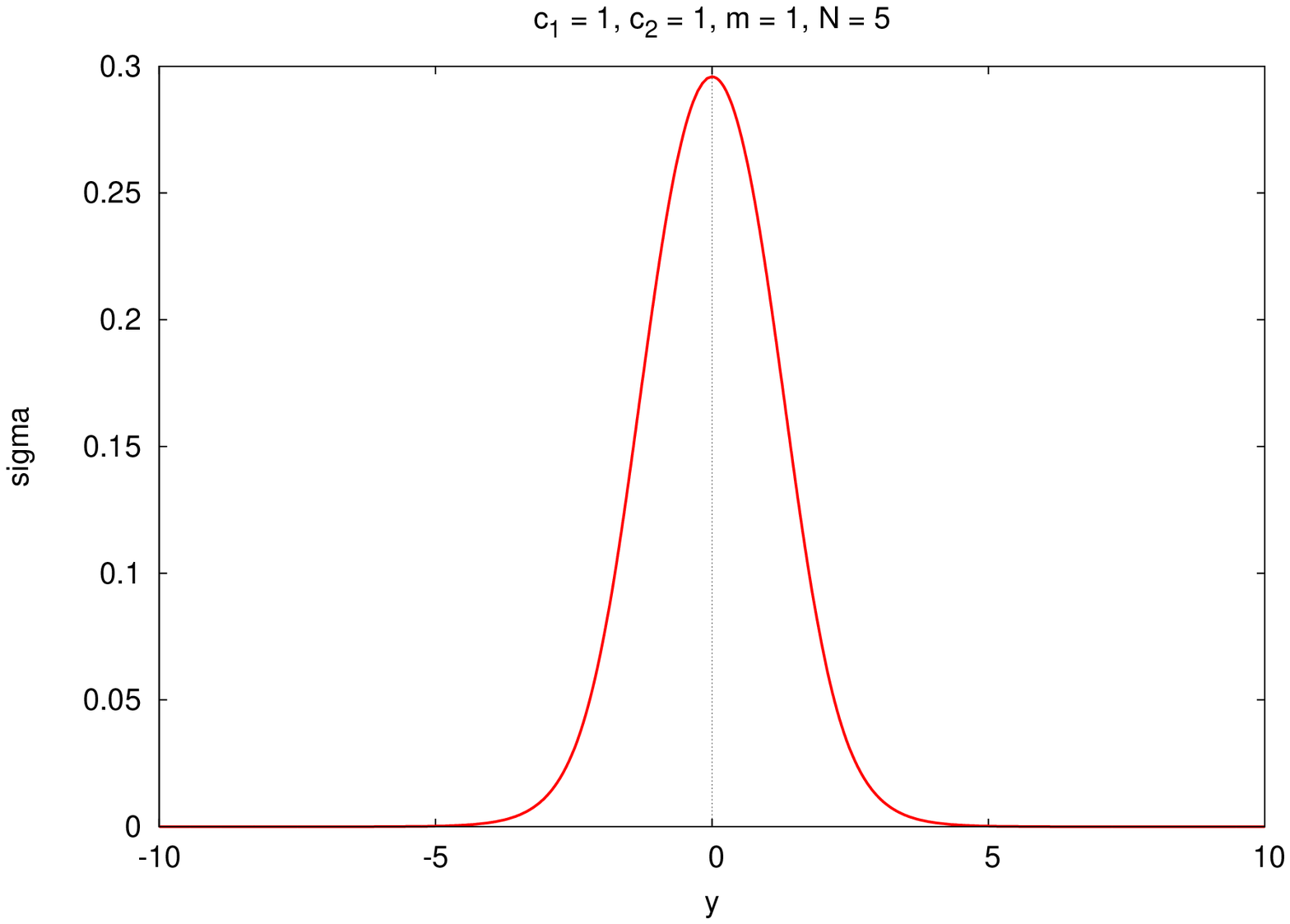}
\end{minipage}
\caption{Profiles of $e^{-\eta}$ and $\sigma$ in the 
coincident case. 
The parameters of the plot are given above the picture. 
Positions of all domain walls are centered at the origin.}
\label{fig:02}
\end{center}
\end{figure}

This set of solutions is not invariant under the 
symmetry transformations \refer{eq:symm1} in general. 
For a choice of $U = \mathbf{1}_N$, however, 
the solutions (\ref{eq:coin1})-(\ref{eq:coin6}) 
 are invariant under the action of the diagonal global 
symmetry $SU(N)_{L+R+c}$. 
We show the $y$-dependence of $e^{-\eta}$ and of 
$\sigma = \partial_y \eta/2$ for the 
coincident case in Fig.\ref{fig:02}.

\begin{figure}
\begin{center}
\begin{minipage}[b]{0.49\linewidth}
\centering
\includegraphics[width=\textwidth]{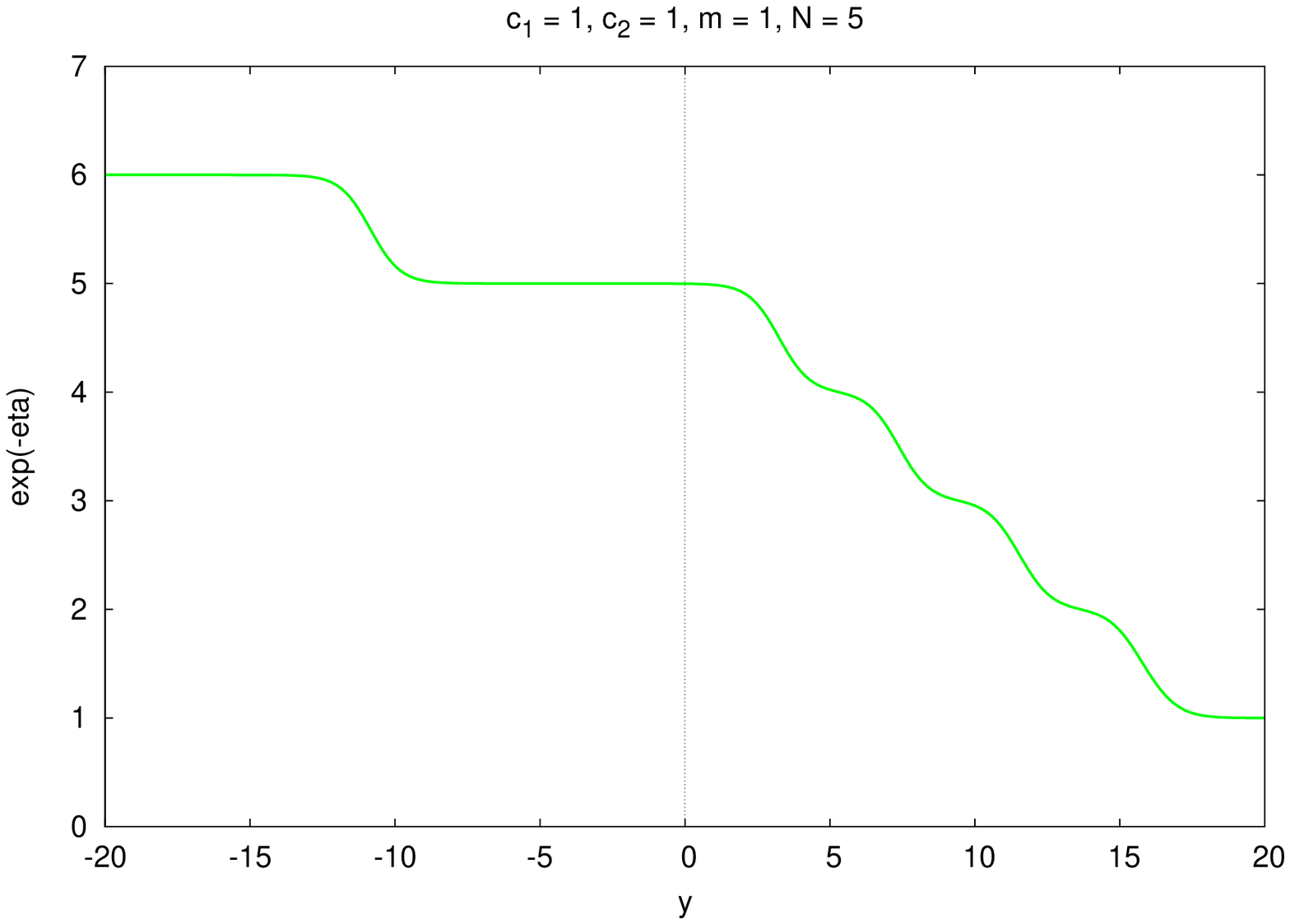}
\end{minipage}
\begin{minipage}[b]{0.49\linewidth}
\centering
\includegraphics[width=\textwidth]{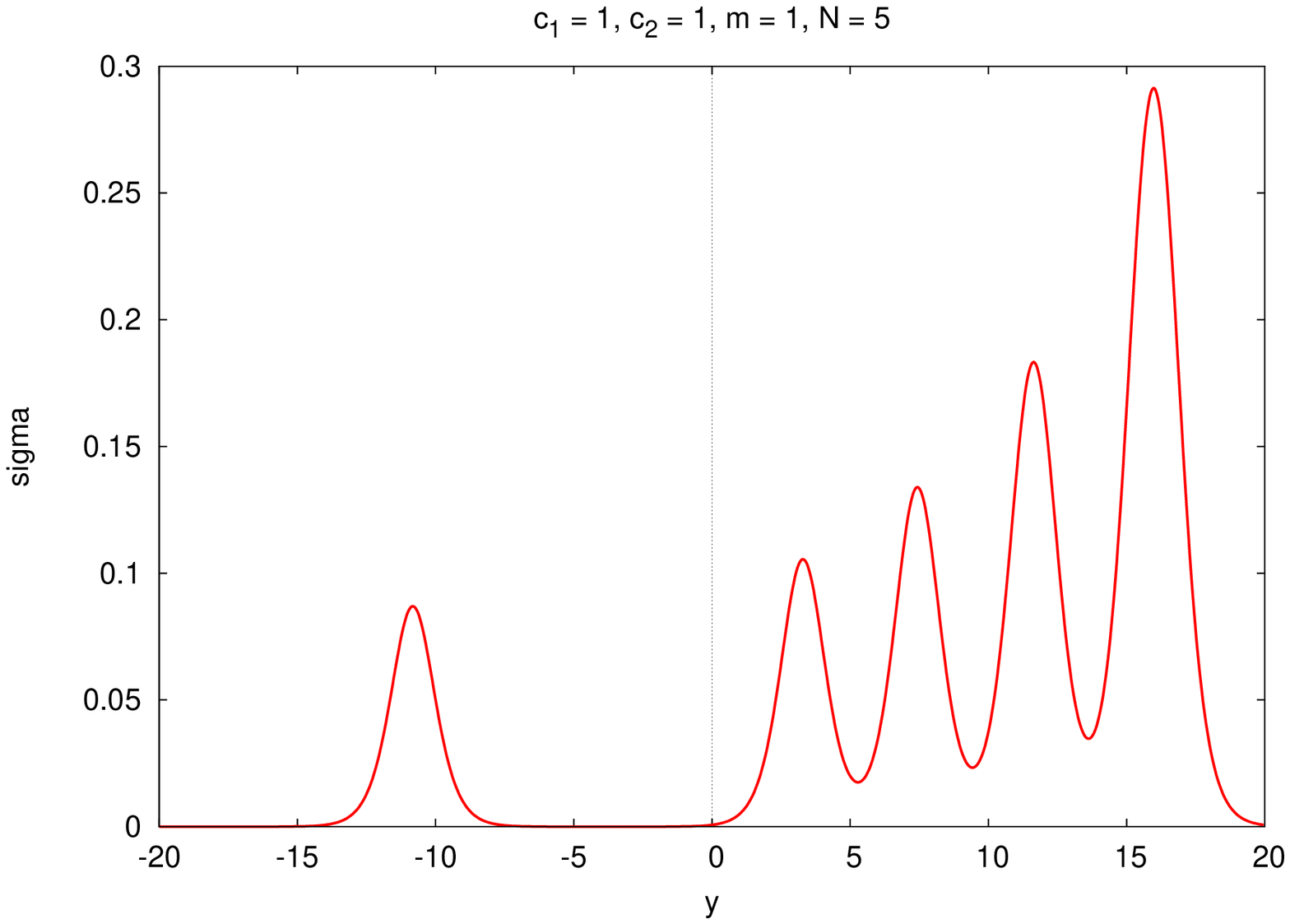}
\end{minipage}
\caption{Profiles of $e^{-\eta}$ and $\sigma$ in the 
non-coincident case. 
The parameters of the plot are given above the picture. 
Positions of domain walls are $y_1 = -10, y_2 = 4, 
y_3 = 8, y_4 = 12$ and $y_5 = 16$.}
\label{fig:03}
\end{center}
\end{figure}

For more general case such as 
non-coincident walls, the dependence of $e^{-\eta}$ and 
$\sigma$ on $y$ is more complicated. 
Furthermore, the equation \refer{eq:x} cannot be solved 
in a closed form in general, except for first few values 
of $N$. 
Thus, one has to use numerical techniques. 
In Fig. \ref{fig:03}, we present an example of five 
non-coincident walls.

\subsection{Localization of non-Abelian gauge fields}

In order to obtain massless gauge fields localized on the 
domain wall we need to introduce a new gauge symmetry 
which is not broken in the bulk.
As we have seen in the previous subsections, the 
coincident domain wall solutions 
(\ref{eq:coin1})-(\ref{eq:coin6}) do not 
break a large part of the global symmetry.
Let us then gauge $SU(N)_{L+R}\equiv SU(N)_V$ and denote 
new gauge fields as $V_{M}$. 
Then the fields $H_1$ and $H_2$ are in the bi-fundamental 
representation of $SU(N)_c\times SU(N)_V$ and the 
covariant derivatives \refer{eq:cov1} are modified to 
\begin{equation}\label{eq:cov2}
\tilde D_M H_{1,2}  = \partial_M H_{1,2} +i W_{M} H_{1,2} -i H_{1,2}V_{M}\,.
\end{equation}
Quantum numbers of the gauged model are summarized in Tab. \ref{table:02}

\begin{table}
\begin{center}
\begin{tabular}{c|cccc|c|c}
\hline
 & $SU(N)_{c}$ & $U(1)_1$ & $U(1)_2$ & $SU(N)_{V}$ 
 & $U(1)_{A}$ & mass\\ \hline
$H_{1}$ & $\square$ & 1 & 0 & $\square$  
& 1 & $m{\bf 1}_{N}$\\ 
$H_{2}$ & $\square$ &  1 & $-1$ & $\square$ 
 & $-1$ & 0 \\ 
$H_{3}$ & {\bf 1} & 0 & 1 & {\bf 1} 
 & 0  & 0\\ 
$\Sigma$ & ${\rm adj}\oplus{\bf 1}$ & 0 & 0 & {\bf 1} 
 & 0 & 0\\
$\sigma$ & ${\bf 1}$ & 0 & 0 & {\bf 1}  & 0 & 0 \\
\hline
\end{tabular}
\end{center}
\caption{Quantum numbers of the gauged  
model.}
\label{table:02}
\end{table}

We introduce field-dependent gauge coupling for $V_M$ as
\begin{equation}\label{eq:pdg}
\frac{1}{2\tilde g^2(\sigma)} = \lambda \sigma\,,
\end{equation}
where we assume that $\lambda$ is a positive constant 
\begin{eqnarray}
\lambda > 0.
\end{eqnarray}
If we denote the field strength for the new gauge fields 
as $\tilde G_{MN}$ the Lagrangian for the gauged model is 
given by 
\begin{equation}\label{eq:gaugedlagr}
\oper{L} = \tilde{\oper{L}}-\frac{1}{2\tilde g^2(\sigma)}
\Tr\Bigl[\tilde G_{MN}\tilde G^{MN}\Bigr]\,,
\end{equation} 
where $\tilde{\oper{L}}$ is the same as in \refer{eq:lag1} 
except that the covariant derivatives \refer{eq:cov1} are 
replaced by \refer{eq:cov2}.
The choice of the field-dependent coupling \refer{eq:pdg} is 
inspired by supersymmetry. 
As discussed in Ref. \cite{Ohta:2010fu}, a term that is 
linear in adjoint scalars, appearing in front of the kinetic 
term for gauge field, naturally arises in five-dimensional 
${\mathcal N}=1$ supersymmetric gauge theories 
\cite{Seiberg:1996bd}. 
In fact, our model can be embedded into ${\mathcal N}=1$ 
supersymmetry.

It is not hard to see that the solution 
(\ref{eq:coin1})-(\ref{eq:coin6}) in the ungauged model 
is equally valid in the gauged model.
If we write down the equation of motion for the new gauge 
field $V_M$ we have 
\begin{equation}
\partial_M \tilde G^{MN} = J^{N}\,,
\label{eq:EOM_gauge_field}
\end{equation}
where $J_M$ stands for the current of $V_M$. 
Since the solution preserves $SU(N)_V$, the current vanishes 
for the domain wall solution 
(\ref{eq:coin1})-(\ref{eq:coin6}), and $V_M = 0$ is a valid 
solution to the equation of motion (\ref{eq:EOM_gauge_field}). 
Then the other equations of motion of the gauged model 
reduce
to those of the ungauged model because of $V_M = 0$. 
Therefore, we see that (\ref{eq:coin1})-(\ref{eq:coin6}) 
in addition to the condition $V_M = 0$ solves the whole set 
of equations of motion of the gauged model.

\subsection{Positivity of position-dependent gauge coupling}

We wish to show that the field-dependent gauge coupling 
\refer{eq:pdg} assures the positive definiteness of the 
position-dependent gauge coupling for any configurations 
of the domain-wall. 
Since we do not need 
the effective theory in full,  
we reserve to derive the rest of the effective Lagrangian  
for the next section. 

For the moment it is sufficient to know, that  
the field-dependent gauge coupling $1/\tilde g^2(\sigma)$ 
is given by its value in the background solution 
\begin{equation}
\frac{1}{\tilde g^2(\sigma)}\bigg|_{\mathrm{background}} 
\equiv  \frac{1}{\tilde g^2(y)} 
= \lambda\partial_y \eta = - \lambda \partial_y \ln x\,,
\end{equation}
where $x = e^{-\eta} \ge 0$ is a solution to \refer{eq:x}. 
Differentiating \refer{eq:x} we find 
\begin{equation}
\frac{1}{x}\partial_y x = - \frac{c_1}{c_2}\sum_{i=1}^N
\frac{e_i }{(1+e_i x)^2}\biggl(2m+\frac{1}{x}
\partial_y x\biggr)\,.
\end{equation}
This leads to the formula
\begin{equation}
\frac{1}{2\tilde g^2(y)} = \frac{\lambda c_1}{c_2}
\sum_{i=1}^N \frac{e_i}{(1+e_ix)^2}\bigg/\biggl(1
+\frac{c_1}{c_2}\sum_{i=1}^N \frac{e_i}{(1+e_ix)^2}\biggr)\,,
\end{equation} 
which is indeed positive in the whole range of 
$y$-coordinate.

Integrating $1/\tilde g^2(y)$ over the extra-dimensional 
coordinate $y$ we obtain the effective gauge coupling in 
$3+1$-dimensional world volume 
\begin{equation}
\frac{1}{\tilde g^2} = \lambda \lineint y\, 
\partial_y \eta 
= \lambda [\eta(\infty)-\eta(-\infty)]\,.
\label{eq:eff_coupling}
\end{equation}
The asymptotic values of $\eta$ are found from 
(\ref{coin1}) as 
\begin{equation}
\eta(\infty) = 0\,, \hspace{0.5cm} \eta(-\infty) 
= -\ln\left(1+N{c_1 \over c_2} \right)\,.
\end{equation}
The
easiest way to see these asymptotic values is to note 
that 
in (\ref{coin2}) we can take the limits of \refer{eq:x} 
to obtain
\begin{equation}
x = 1 + \frac{c_1}{c_2}\sum_{i=1}^N \frac{1}{1+e_i x} 
\longrightarrow 
\Bigg\{\begin{array}{lcr}
1 & \mathrm{if} & y \to \infty\,, \\
1+\tfrac{Nc_1}{c_2} & \mathrm{if} & y \to -\infty\,,
\end{array}
\end{equation}
since $\eta$ is finite  at both infinities 
\footnote{
We can just look at \refer{eq:h03} with 
$H_3^0 = \sqrt{c_2}$ and recall \refer{eq:vev} 
to obtain the same result.
}.

Thus, the effective gauge coupling is given as 
\begin{equation}\label{eq:effcoupl2}
\frac{1}{\tilde g^2} = \lambda \ln\biggl(1
+N\frac{c_1}{c_2}\biggr)\,.
\end{equation}
It is interesting to observe that the 
effective gauge coupling is proportional to the width 
of the domain wall $\ln(1+N{c_1}/{c_2})$, 
which is not a modulus, but is 
fixed by parameters of the theory. 
This feature is in sharp contrast to that in 
Ref.~\cite{Arai:2012cx} where the effective gauge 
coupling constant is proportional to the domain wall 
width, which is a modulus undetermined by the theory. 
We have now confirmed the stability of the gauge kinetic 
term (by choosing the parameters of the theory as 
$\lambda, m, c_1,c_2 > 0$).
\begin{figure}
\begin{center}
\begin{minipage}[b]{0.45\linewidth}
\centering
\includegraphics[width=\textwidth]{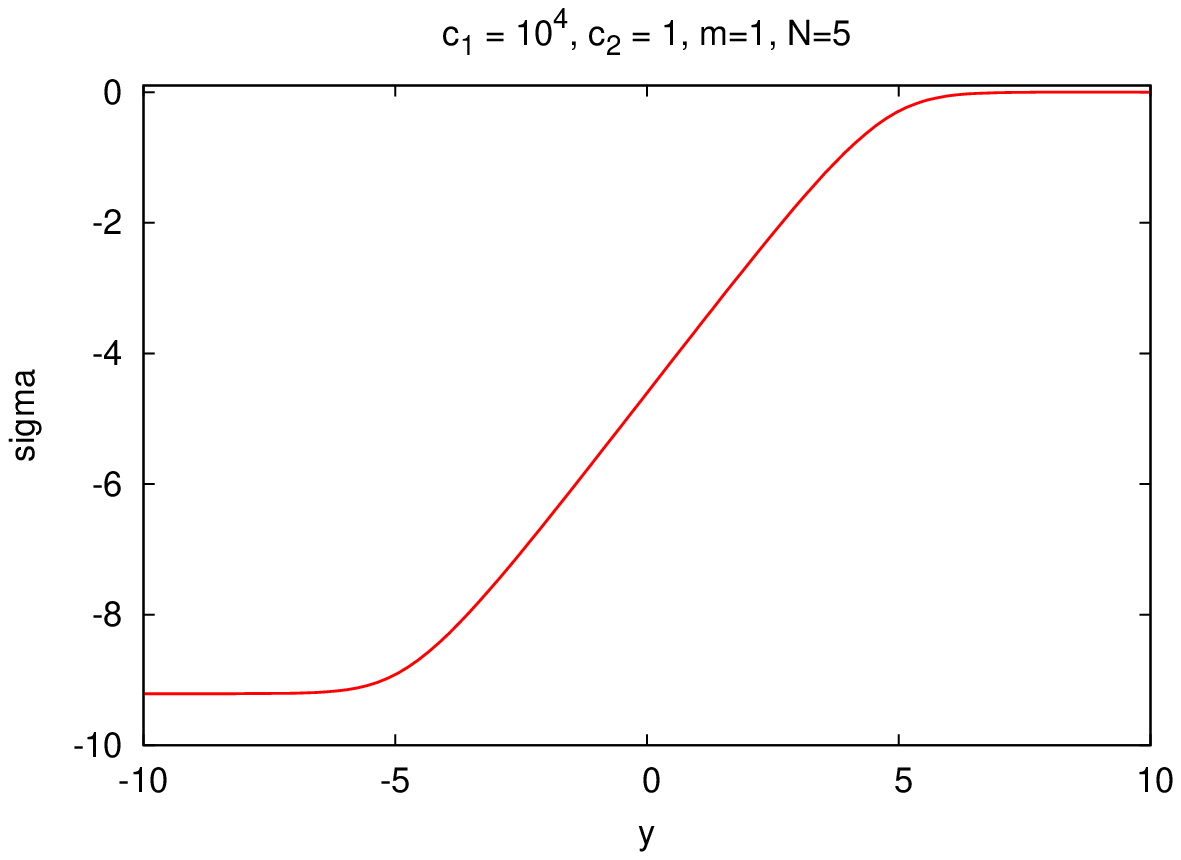}
\end{minipage}
\begin{minipage}[b]{0.45\linewidth}
\centering
\includegraphics[width=\textwidth]{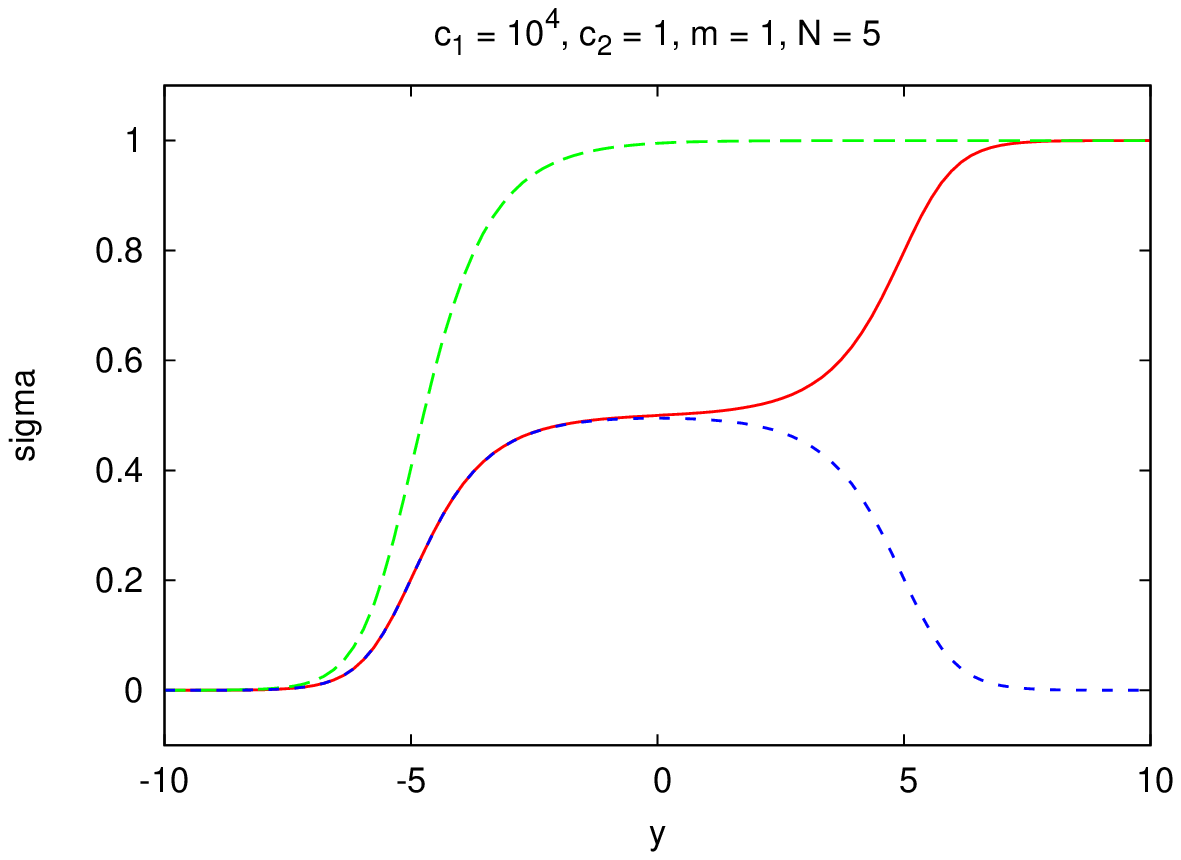}
\end{minipage}
\caption{Profile of $\eta$-kink is shown in the left 
panel for  the coincident case.
In the right panel, 
plots of $\Tr[\Sigma]$ (green dashed curve),
$\Tr[\Sigma]-\sigma$ (red solid curve), and
$\sigma$ (blue dotted curve) are shown.
}
\label{fig:eta}
\end{center}
\end{figure}

Eq.~(\ref{eq:eff_coupling}) shows that the effective 
four-dimensional gauge coupling constant $\tilde g$ is 
determined only by the boundary conditions at infinity. 
It can be interpreted as a kind of topological charge 
of the $\eta$-kink, whose profile is shown 
in Fig.~\ref{fig:eta}.

Note that we have considered only the BPS solutions so 
far, and confirmed the stability of the model.
The anti-BPS solutions are also stable, since 
they can are obtained by the parity transformation 
$x \to - x$.

\section{Effective Lagrangian}\label{sc:efftheory}

In this section we calculate the low-energy effective 
Lagrangian on the background domain wall 
solution
in the moduli approximation \cite{Ma}, where the moduli 
are promoted to fields on the world-volume with 
coordinates $x^\mu$ and 
are
assumed to depend only weakly 
on $x^\mu$. 
As the background we consider the coincident walls 
$\phi_{\tiny bg}=my_0 {\bf 1}_N$. 
Our model have two moduli: a Hermitian $N\times N$ 
matrix $\phi$ and a unitary $N\times N$ matrix $U$. 
We study in two steps: taking only $U$ as moduli fields, 
and then both $\phi$ as well as $U$.

\subsection{Effective Lagrangian for chiral fields 
} 
Here we ignore the moduli fields 
$\phi$, 
and take only $U$ moduli fields, that 
describe the Nambu-Goldstone modes 
associated with the symmetry breaking in 
Eq.(\ref{eq:nambu_goldstone_mode}). 
When $\phi=my_0 {\bf 1}_N$, the solution 
\refer{eq:h01}-\refer{eq:h03} with \refer{eq:ms} 
and \refer{eq:decomp} become in the strong gauge 
coupling limit $g\rightarrow \infty$ 
and $e\rightarrow \infty$ as 
\begin{align}
H_1 & = \sqrt{c_1}e^{my}\Omega^{-1/2}\,, \label{sol1} \\
H_2 & = \sqrt{c_1}e^{\eta/2}e^{my_0(x)}
\Omega^{-1/2}U^{\dagger}(x)\,, \label{sol2} \\
H_3 & = \sqrt{c_2}e^{-\eta/2}\,,\label{sol3}
\end{align} 
where $\Omega$ and $\eta$ are defined in 
Eqs.(\ref{coin1}) and (\ref{coin2}) with the replacement 
$y_0 \to y_0(x^\mu)$ and $U\rightarrow U(x^\mu)$. 
We plug these 
into the Lagrangian (\ref{eq:gaugedlagr}),  
where the gauge fields $W_\mu$ and $A_{\mu}$ are no longer 
dynamical and should be eliminated as auxiliary fields. 
After integrating over the extra-dimensional coordinate and 
taking up to quadratic terms in the derivatives, 
we obtain the low energy effective Lagrangian (the 
detailed calculation is given in Appendix \ref{appendix}) 
\begin{align}
\oper{L}_{\rm eff} & = \frac{c_1}{2m}
\biggl[ (\alpha+1)\Tr\bigl[\oper{D}_{\mu}U^{\dagger}
\oper{D}^{\mu}U\bigr] + 
\frac{\alpha}{N}\, \Tr\bigl[U\oper{D}_{\mu}U^{\dagger}
\bigr]\Tr\bigl[U\oper{D}^{\mu}U^{\dagger}\bigr]
\biggr]\nonumber \\\label{result} 
& +\frac{Nmc_1}{2}\partial_{\mu}y_0\partial^{\mu}y_0
-\frac{1}{2m}\ln\biggl(1+\frac{Nc_1}{c_2}\biggr)
\Tr\Bigl[\tilde G_{\mu\nu}\tilde G^{\mu\nu}\Bigr]\,,
\end{align}
where 
\begin{equation}
{\mathcal D}_\mu U =\partial_\mu U+i[V_\mu,U],
\end{equation}
and
\begin{equation}
\alpha  \equiv \frac{1}{2}+\frac{c_2}{Nc_1}-\frac{c_2}{Nc_1}\left(1+\frac{c_2}{Nc_1}\right)\ln \biggl(1+\frac{Nc_1}{c_2}\biggr)\,.
\label{eq:alpha}
\end{equation}

Let us define the decay constants $f_\pi$ for the adjoint 
field and $f_\eta$ for the singlet field 
\begin{equation}
f_{\pi} = \sqrt{\frac{c_1(\alpha+1)}{2m}}, \qquad
f_{\eta} = \sqrt{\frac{c_1}{Nm}}\,.
\end{equation}
Then the canonically normalized adjoint and singlet 
fields can be defined as 
$\hat \pi$ and $\eta$ respectively 
\begin{equation}
\frac{1}{f_\pi}\oper{D}_{\mu}\hat \pi 
= i \Bigl[U\oper{D}_{\mu}U^{\dagger}
-\frac{\mathbf{1}_N}{N}\Tr\Bigl(U\oper{D}_{\mu}U^{\dagger}\Bigr)\Bigr],
\end{equation}
\begin{equation}\label{eta1}
\frac{1}{f_\eta}\partial_{\mu}\eta :
= i \Tr\Bigl(U\oper{D}_{\mu}U^{\dagger}\Bigr)\,.
\end{equation}
The effective Lagrangian Eq. \refer{result} can be 
rewritten as 
\begin{equation}\label{eta2}
\oper{L}_{\rm eff} = \Tr\Bigl(\oper{D}_{\mu}\hat\pi\oper{D}^{\mu}\hat\pi\Bigr)+\frac{1}{2}\partial_{\mu}\eta\partial^{\mu}\eta
+\frac{Nmc_1}{2}\partial_{\mu}y_0\partial^{\mu}y_0-\frac{1}{2\tilde g^2}\Tr\Bigl[\tilde G_{\mu\nu}\tilde
G^{\mu\nu}\Bigr]\,. 
\end{equation}
A similar effective Lagrangian has been obtained in 
studying Skyrmions realized inside domain 
walls \cite{Eto:2005cc}. 
A new feature of  
\refer{eta2}
compared with our 
previous work \cite{Arai:2012cx} is that the coupling 
strength $f_\pi$ of the adjoint field is larger 
\footnote{
Difference of factors of $2$ and $N$ is due to a 
convention of $SU(N)$ matrix normalization 
$\Tr(T^I T^J)=\delta^{IJ}/2$. 
} than 
$f_\eta$ of the singlet field by 
a factor $\sqrt{\alpha+1}$.

\subsection{Full effective Lagrangian}

As the final step, we consider the general case of both 
$U$ and $\phi$ as moduli fields. 
Then scalar fields are given in terms of moduli fields 
 $U(x^\mu)$ and $\phi(x^\mu)$ as \footnote{
The center of mass position of walls $y_0(x^\mu)$ 
is a free field as in Eq.(\ref{eta2}) and is 
suppressed here. 
}
\begin{align}
H_1 & = \sqrt{c_1}e^{my}\Omega^{-1/2}\,, 
\label{Higgs-g1}\\
H_2 & = \sqrt{c_1}e^{\eta/2}\Omega^{-1/2}
e^{\phi}U^{\dagger}\,, \label{Higgs-g2}\\
H_3 & = \sqrt{c_2}e^{-\eta/2}\,, \label{Higgs-g3}
\end{align} 
\begin{align}
\Omega^{-1} & = \frac{e^{-\eta}e^{-2\phi}}
{\mathbf{1}_N+e^{2my}e^{-2\phi}e^{-\eta}}\,, \\
\label{eq:geneta} e^{-\eta} & = 1
+\frac{c_1}{c_2}\Tr\left(\frac{\mathbf{1}_N}
{\mathbf{1}_N+e^{2my}e^{-2\phi}e^{-\eta}}\right)\,.
\end{align}

To obtain the effective Lagrangian, we need to repeat 
the same procedure as in the previous subsection, 
where the covariant derivatives acting on functions of 
matrices require more care (see e.g. 
Appendix B
in Ref. \cite{Arai:2012cx}) and 
causes difficulty when deriving the closed form of 
the effective Lagrangian. 
However, we have a convenient parameter to expand the 
effective Lagrangian, the ratio $c_1/c_2$ whose 
logarithm has a physical meaning as the width of 
the domain wall (see Eq.(\ref{eq:effcoupl2})). 
As given in Appendix \ref{appendix}, the effective 
Lagrangian up to the order of $c_1/c_2$ is given as 
\begin{equation}\label{eq:efflagr}
\oper{L}_{\rm eff} = \oper{L}_{\rm eff}^{(0)}
+\oper{T}_{\phi}^{(1)}+\oper{T}_{U}^{(1)}
+\oper{T}_{mix}^{(1)}+\oper{T}_{\phi}^{\prime}
+\oper{T}_{U}^{\prime}
+c_1O\Bigl((c_1/c_2)^2\Bigr)\,,
\end{equation}
where 
\begin{align}
 \oper{L}_{{\rm eff}}^{(0)} & = \frac{c_1}{2m}
\Tr\biggl[\oper{D}_{\mu} \phi\,
 \frac{\cosh(\oper{L}_{ \phi})-1}{\oper{L}_{ \phi}^2
\sinh(\oper{L}_{ \phi})}
\ln\biggl(\frac{1+\tanh(\oper{L}_{ \phi})}
{1-\tanh(\oper{L}_{ \phi})}\biggr)(\oper{D}^{\mu}\phi) 
\nonumber \\
& +U^{\dagger}\oper{D}_{\mu}U\,
\frac{\cosh(\oper{L}_{\phi})-1}{\oper{L}_{ \phi}
\sinh(\oper{L}_{ \phi})}
\ln\biggl(\frac{1+\tanh(\oper{L}_{ \phi})}
{1-\tanh(\oper{L}_{ \phi})}\biggr)(\oper{D}^{\mu} \phi) 
\nonumber \\\label{eq:efflagr0}
&+\frac{1}{2}\oper{D}_{\mu}U^{\dagger}U\frac{1}
{\tanh(\oper{L}_{\phi})}\ln\biggl(
\frac{1+\tanh(\oper{L}_{\phi})}{1-\tanh(\oper{L}_{ \phi})}
\biggr)
(U^{\dagger}\oper{D}^{\mu}U)\biggr]\,,
\end{align}
with a Lie derivative with respect to $A$, 
${\mathcal L}_A(B)=[A,B]$.
Interestingly, this is the same effective Lagrangian we 
obtained previously \cite{Arai:2012cx}. 
The rest of terms are of order of $c_1/c_2$, given by 
\begin{align}
\oper{T}_{U}^{(1)} & = \frac{c_1^2}{16c_2m}
\Tr\Bigl[\oper{D}_{\mu}U^{\dagger}UF_U(\partial_x,
\oper{L}_{\phi})(U^{\dagger}\oper{D}^{\mu}U)
e^{x\phi}\Bigr]\Tr\Bigl[e^{-x\phi}\Bigr]\bigg|_{x=0}\,, \\
\oper{T}_{mix}^{(1)} & 
=  \frac{c_1^2}{8c_2m}\Tr\Bigl[U^{\dagger}
\oper{D}_{\mu}UF_{mix}(\partial_x,\oper{L}_{\phi})
(\oper{D}^{\mu}\phi)e^{x\phi}\Bigr]
\Tr\Bigl[e^{-x\phi}\Bigr]\bigg|_{x=0}\,, \\
\oper{T}_{\phi}^{(1)} & =  \frac{c_1^2}{8c_2m}
\Tr\Bigl[\oper{D}_{\mu}\phi F_{\phi}
(\partial_x,\oper{L}_{\phi})(\oper{D}^{\mu}\phi)
e^{x\phi}\Bigr]
\Tr\Bigl[e^{-x\phi}\Bigr]\bigg|_{x=0}\,, \\
\oper{T}_{\phi}^{\prime} & = -\frac{c_1^2}{16c_2m}
F(\partial_x)\Tr\Bigl[e^{x\phi}\oper{D}_{\mu}\phi\Bigr]
\Tr\Bigl[e^{-x\phi}\oper{D}^{\mu}\phi\Bigr]\bigg|_{x=0}\,, 
\\
\oper{T}_{U}^{\prime} & = \frac{c_1^2}{16c_2m}
F(\partial_x)\Tr\Bigl[e^{x\phi}\oper{D}_{\mu}U^{\dagger}
U\Bigr]
\Tr\Bigl[e^{-x\phi}\oper{D}^{\mu}U^{\dagger}U\Bigr]
\bigg|_{x=0},
\end{align}
and 
\begin{align}
F_U(x,\oper{L}_{\phi}) & = \lineint y\, 
\frac{\cosh(\oper{L}_{\phi})}{\cosh^2(y-x)\cosh(y)
\cosh(y-\oper{L}_{\phi})}\,, \\
F_{mix}(x,\oper{L}_{\phi})  & = \lineint y\, 
\frac{\cosh(y)\cosh(y-\oper{L}_{\phi})
-\sinh(y)\sinh(y-\oper{L}_{\phi})-1}
{\oper{L}_{\phi}\cosh(y)\cosh(y-\oper{L}_{\phi})
\cosh^2(y-x)}\,, \\
F_{\phi}(x,\oper{L}_{\phi}) & = \lineint y\, 
\frac{\cosh(y)\cosh(y+\oper{L}_{\phi})
-\sinh(y)\sinh(y+\oper{L}_{\phi})-1}
{\oper{L}_{\phi}^2\cosh(y)\cosh(y+\oper{L}_{\phi})
\cosh^2(y-x)}\,, \\
F(x) & = \lineint y\, \frac{1}{\cosh^2(y)\cosh^2(y-x)}\,.
\end{align}
All the above integrals can be obtained in closed 
forms.

\section{More general models of stable 
position-dependent coupling}

In this section, we wish to show that there are more 
models with the stable position-dependent coupling. 
We will illustrate the point by extending the model 
to include more fields and more gauge symmetry. 

The position-dependent gauge coupling comes from the 
cubic coupling between a singlet scalar field and 
field strengths of non-Abelian gauge fields in $4+1$ 
dimensions 
\begin{eqnarray}
{\mathcal L}_{\rm cubic} = 
C(\sigma_i) \Tr\left[\tilde G_{MN}\tilde G^{MN}\right], 
\end{eqnarray}
where the function $C(\sigma_i)$ of singlet scalar fields 
$\sigma_i$ should be linear, if it is to be embeddable 
into a supersymmetric gauge theory in $4+1$ dimensions 
\begin{eqnarray}
C(\sigma_i) = \sum_i \gamma_i \sigma_i,
\quad \gamma_i \in \mathbb{R}, 
\label{eq:coefficient_singlets}
\end{eqnarray}
where $\gamma_i$ are constant coefficients.

Usually each domain wall has one complex moduli: 
a position and a phase. 
For example, both the massive $\mathbb{C}P^2$ model 
and the massive 
$\mathbb{C}P^1 \times \mathbb{C}P^1$ model 
have two free domain walls, corresponding to the 
two complex moduli. 
Although two free domain walls can produce the desired 
profile of position-dependent gauge coupling by an 
appropriate choice of parameters in 
Eq.(\ref{eq:coefficient_singlets}), they provide the 
undesired modulus for the width of the profile. 
To avoid this problem, we are led to consider models 
with a single complex moduli. 
The simplest one of such models is the three flavor model 
in Ref.\cite{Ohta:2010fu}, where three scalar fields are 
constrained by the two Abelian gauge symmetry 
$U(1)\times U(1)$. 
Geometry of this three-flavor model is examined in 
Appendix \ref{ap:3flavor}. 
Our model in this paper is an extension of this model 
to non-Abelian gauge group: 
$U(1)\times U(1) \to U(N)\times U(1)$. 
The next-simplest possibility is to consider four scalars 
constrained by three Abelian gauge symmetry 
$U(1)\times U(1)\times U(1)$, which we call 
four-flavor models. 
We give quantum numbers of fields of a typical four-flavor 
model in Tab.\ref{table:04}. 
In the limit of strong gauge couplings, the gauge theory 
becomes a nonlinear sigma model whose target space is 
given by an intersection of three conditions as 
\begin{eqnarray}
\left(\mathbb{C}^2 \times \mathbb{C}P^1\right)
\cap
\left(\mathbb{H}^2 \times \mathbb{C}P^1\right)
\cap
\left(\mathbb{C}^2 \times \mathbb{C}P^1\right)
 \simeq \mathbb{C}P^1.
\end{eqnarray}

\begin{table}
\begin{center}
\begin{tabular}{c|ccc|c}
\hline
 & $U(1)_1$ & $U(1)_2$ & $U(1)_3$ & mass\\ \hline
$H_{1}$ & 1 & 0 & 0& $m_1$\\ 
$H_{2}$ & 1 & $-1$ & 0& $m_2$\\ 
$H_{3}$ &  0 & 1 & 1& $m_3$\\ 
$H_{4}$ &  0 & 0 & 1& $m_4$\\ 
$\sigma_1$ & 0 & 0 & 0& 0\\
$\sigma_2$ & 0 & 0 & 0& 0 \\
$\sigma_3$ & 0 & 0 & 0& 0 \\
\hline
\end{tabular}
\end{center}
\caption{Quantum numbers of the $U(1)_1\times U(1)_2
\times U(1)_3$ four-flavor model.}
\label{table:04}
\end{table}

The vacuum condition is given by 
\begin{eqnarray}
|H_1|^2 + |H_2|^2 =  c_1,\quad
- |H_2|^2 + |H_3|^2 = c_2,\quad
|H_3|^2 + |H_4|^2 = c_3,\\
H_1(\sigma_1 - m_1) = 0,\quad
H_2(\sigma_1 - \sigma_2 - m_2) = 0,\qquad\qquad\\
H_3(\sigma_2 + \sigma_3 - m_3) = 0,\quad
H_4(\sigma_3 - m_4) = 0,\qquad\qquad
\end{eqnarray}
where $c_i$ is the Fayet-Iliopoulos parameter of $U(1)_i$ and $m_a$ is 
the mass for $H_i$. 
All possible solutions to these equations are shown 
in Tab.~\ref{table:05}.
There are four solutions but only two of them are 
valid solutions for any choice of real parameters 
of $c_i$. 
When we choose $c_1>0$, $c_2>0$ and $c_3 > c_1 + c_2$, 
we are left with the vacua $\left<1\right>$ and 
$\left<2\right>$ in Tab.~\ref{table:05}.
\begin{table}
\begin{center}
\begin{tabular}{c|ccccccc}
 & $|H_1|$ & $|H_2|$ & $|H_3|$ & $|H_4|$ & $\sigma_1$ 
& $\sigma_2$ & $\sigma_3$\\
\hline
$\left<1\right>$ & $0$ & $\sqrt{c_1}$ & $\sqrt{c_{1+2}}$ 
& $\sqrt{c_{3-1-2}}$ & $m_2+m_3-m_4$ & $m_3-m_4$ & $m_4$\\
$\left<2\right>$ & $\sqrt{c_1}$ & $0$ & $\sqrt{c_2}$ 
& $\sqrt{c_{3-2}}$ & $m_1$ & $m_3-m_4$ & $m_4$\\
$\left<3\right>$ & $\sqrt{c_{1+2}}$ & $\sqrt{-c_2}$ 
& $0$ & $\sqrt{c_3}$ & $m_1$ & $m_1-m_2$ & $m_4$\\
$\left<4\right>$ & $\sqrt{c_{1+2-3}}$ & $\sqrt{c_{3-2}}$ 
& $\sqrt{c_3}$ & $0$ & $m_1$ & $m_1-m_2$ & $-m_1+m_2+m_3$
\end{tabular}
\caption{VEVs of candidate vacua: 
We use abbreviations like 
$c_{3-1-2} \equiv c_3-c_1-c_2$.}
\label{table:05}
\end{center}
\end{table}

The moduli matrix formalism \cite{Eto:2006pg,Eto:2005wf} 
is powerful enough to give generic solutions of the 
BPS equations of this nonlinear sigma model. 
Especially, we are 
interested in the kink profiles of $\sigma_1$, $\sigma_2$ 
and $\sigma_3$. They 
can be expressed as derivatives of real functions 
$\eta_i$ 
\begin{eqnarray}
\sigma_i = \frac{1}{2}\partial_y \eta_i,\quad (i=1,2,3),
\end{eqnarray}
where the real functions $\eta_i$ are determined by 
the following algebraic conditions 
\begin{eqnarray}
e^{-\eta_1+2m_1y}+e^{-\eta_1+\eta_2+2m_2y-2a}=c_1,\\
-e^{-\eta_1+\eta_2+2m_2y-2a}+e^{-\eta_2-\eta_3+2m_3y}=c_2,\\
e^{-\eta_2-\eta_3+2m_3y}+e^{-\eta_3+2m_4y}=c_3,
\end{eqnarray} 
with a real constant $a$. 
The parameter $a$ is (the real part of) the unique modulus 
of the solution, corresponding to the position of the domain 
wall.
As we expected, we find only a single modulus. 
The width of the domain wall is not a modulus, but 
fixed by the theory.  

\begin{figure}
\begin{center}
\begin{minipage}[b]{0.48\linewidth}
\centering
\includegraphics[width=\textwidth]{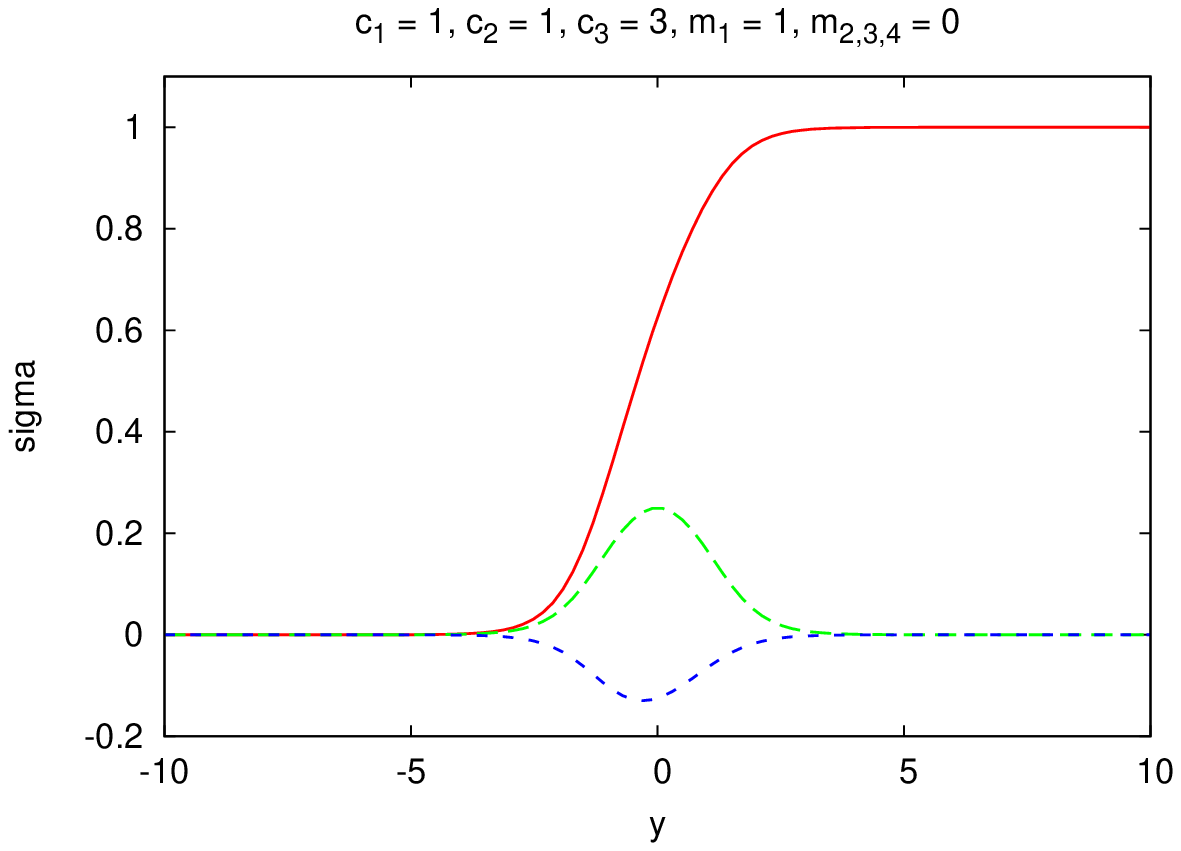}
\end{minipage}
\begin{minipage}[b]{0.48\linewidth}
\centering
\includegraphics[width=\textwidth]{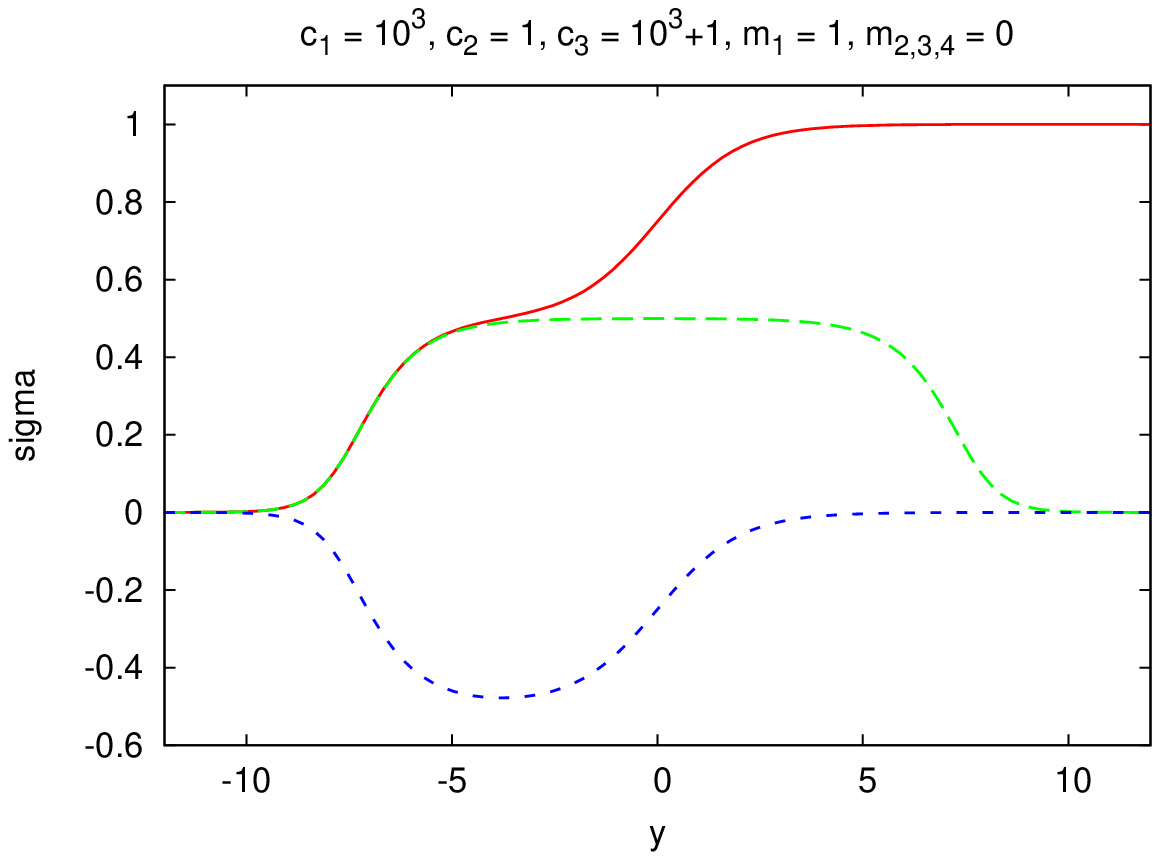}
\end{minipage}
\caption{The kink profiles of $\sigma_1$ (red solid line
), 
$\sigma_2$ (green dashed line
) and $\sigma_3$ (blue dotted line
) for 
two different sets of model parameters. }
\label{fig:sigma_extended}
\end{center}
\end{figure}

Since we want a configuration that $\sigma_i \to 0$ at 
both spacial infinities, we can choose $m_3=m_4=0$. 
Then $\sigma_2 = \sigma_3 \to 0$ at $y=\pm \infty$. 
Several numerical solutions are displayed in 
Fig.~\ref{fig:sigma_extended}.
From Fig.~\ref{fig:sigma_extended}, we clearly see 
that $\sigma_2 \ge 0$ and $\sigma_3 \le 0$ for all the 
values of $y$.
As a position-dependent gauge coupling, we can choose a 
two-parameter family of desirable models 
\begin{eqnarray}
{\mathcal L}_{\rm cubic} = - \left(\gamma_2 \sigma_2 
- \gamma_3\sigma_3\right)\Tr\left[\tilde G_{MN}
\tilde G^{MN}\right],
\end{eqnarray}
where $\gamma_{2,3}$ can be any non-negative real numbers. 
This class of models can be easily made to localize 
non-Abelian gauge fields and minimally interacting 
matter fields by extending two of the $U(1)$ factor 
groups to (possibly different) $U(N)$ gauge groups 
with $N$ scalar fields in the fundamental 
representations, similarly to our model in previous 
sections. 
A new interesting feature of the four-flavor model 
(and its non-Abelian extensions) is that two possible 
profile of singlet fields $\sigma_1, \sigma_2$ can 
provide a different profile for different non-Abelian 
gauge groups such as $SU(3)$, $SU(2)$ and $U(1)$ 
with their associated matter fields.

\section{Conclusion and discussion}

We have successfully stabilized gauge fields and matter fields 
that are localized on domain walls. 
The low-energy effective Lagrangian on the domain wall has 
been worked out, where the adjoint matter fields is 
found
to couple more strongly than singlet matter fields. 
We also explored possible generalization of our stabilization 
mechanism by including more fields and more gauge  
symmetries
and found a class of more generic models with an added 
flexibility  
for
model building with different localization 
profile for different gauge groups.

To build realistic models of brane-world with our 
scenario of localized gauge fields and matter
fields, 
we should address several questions. 
Perhaps the most important question is to obtain 
(massless) matter fields in representations like the 
fundamental rather than the adjoint of localized gauge 
fields. 
One immediate possibility is to use the localization 
mechanism of fermions in a kink background. 
It has been found that zero modes of such fermions are 
localized in such a way to give automatically chiral 
fermions \cite{RuSh}. 
We will pursue this direction and associated anomaly 
questions further. 

Secondly, we should devise a way to give small masses 
to our matter fields in order to do phenomenology. 
Since some of our matter fields are the Nambu-Goldstone 
modes of a broken global symmetry, we need to consider 
an explicit breaking of such global symmetry.

Thirdly, another question is to study possibility of 
supersymmetric model of gauge field localization. 
We need to settle the issues of possible new moduli 
in that case. 
Moreover, we should examine the mechanism of supersymmetry 
breaking. 

Another interesting possibility for model building is 
to localize gauge fields of different gauge groups with 
different profiles. 
This situation is often proposed in recent brane-world 
phenomenology, for instance in Ref.\cite{Hosotani:2008by}. 

Finally, let us examine similarities and differences 
of our domain walls compared to D-branes. 
The most interesting similarity of our domain wall 
with D-branes is the realization of geometrical 
Higgs mechanism, where massless gauge fields in the 
coincident wall become massive as walls separate. 
Detailed analysis of this phenomenon is worth doing. 
On the other hand, 
there are differences as well. 
D-branes in string theory are defined by the Dirichlet 
condition for fundamental string attached to it,  
but no such condition is visible in our domain walls. 
Domain walls similar to ours have been constructed 
in Ref.\cite{Shifman:2002jm}, 
where a probe magnetic charge is placed in 
the bulk Higgs phase. 
The magnetic flux from the probe magnetic monopole is 
carried by a vortex which can end on the domain wall. 
They observed that this wall-string junction 
configuration resembles an open string ending on D-branes 
in string theory. 
However, this phenomenon is in theories with one 
dimension less, namely the world volume of domain walls 
has only $1+2$ dimensions (fundamental theory is 
in $1+3$ dimensions). 
In our model based on a theory in $1+4$ dimensions, 
monopole (codimension three) is a string-like soliton, 
and might possibly be a candidate of something 
similar to the fundamental string. 
It is worth pursueing a possibility of composite solitons 
consisting of domain walls and other solitons such as 
monopoles, in order to clarify more similarities 
of solitons with D-branes. 
Another interesting issue of the effective action on 
D-branes is the non-Abelian generalization of the 
Dirac-Born-Infeld (DBI) action. There have been studies to 
obtain first few corrections to Yang-Mills action in 
string theories. 
Since we are at present interested in up to 
quadratic terms in derivatives, we have found the ordinary 
quadratic action of Yang-Mills fields together with 
the action of moduli fields interacting minimally with 
the Yang-Mills 
fields in addition to the nonlinear interactions among 
themselves. Our result is trivially consistent with the 
quadratic approximation of DBI action, but does not give 
us informations on non-Abelian generalization of DBI 
action. 
In order to shed lights on that issue, 
we need to compute higher 
derivative corrections to our effective Lagrangian, 
which is an interesting future problem.

\section*{Acknowledgements} 
This work is supported in part by Japan Society for the 
Promotion of Science (JSPS) and Academy of Sciences of 
the Czech Republic (ASCR) under the Japan - Czech Republic 
Research Cooperative Program, and by Grant-in-Aid for 
Scientific Research from the Ministry of Education, Culture, 
Sports, Science and Technology, Japan No.21540279 (N.S.), 
No.21244036 (N.S.), and No.23740226 (M.E.). 
The work of M.A.~and F.B.~is supported in part by the 
Research Program MSM6840770029 and by the project of 
International Cooperation ATLAS-CERN of the Ministry of 
Education, Youth and Sports of the Czech Republic.

\appendix
\section{Derivation of the effective Lagrangian}
\label{appendix}
In this appendix, we derive the effective Lagrangians 
\refer{result} and \refer{eq:efflagr} in the strong coupling 
limit. 
Since we are interested in the low energy effective 
Lagrangian, we focus on up the quadratic terms in 
derivatives, which we call ${\mathcal L}^{(2)}$.

\subsection{Eliminating gauge fields}

Here we eliminate gauge fields $W_\mu, A_\mu$ to obtain 
a nonlinear sigma model. 
Starting with $\tilde{\mathcal L}$ in \refer{eq:gaugedlagr} 
in the strong gauge coupling limit $g\rightarrow \infty$ 
and $e \rightarrow \infty$, ${\mathcal L}^{(2)}$ 
is given as
\begin{equation}\label{eff}
\oper{L}^{(2)} = \Tr\abs{D_{\mu}H_1}^2+\Tr\abs{
(D_\mu-i A_\mu)H_2}^2+\abs{(\partial_\mu+i A_\mu)H_3}^2\,,
\end{equation} 
where the covariant derivatives are given as:
\begin{equation}
D_{\mu}H_{1,2} = \hat D_{\mu}H_{1,2}+i W_{\mu}H_{1,2} 
= \partial_{\mu}H_{1,2}-i H_{1,2}V_{\mu}+i W_{\mu}H_{1,2}\,.
\end{equation}
Here we singled-out covariant derivatives $\hat D_{\mu}$ 
containing $V_{\mu}$ fields associated with 
the gauged part of 
the
flavor symmetry $SU(N)_{L+R}$.
Notice that $W_\mu$ and $A_{\mu}$ are no longer dynamical, 
but they are merely auxiliary fields to be eliminated.
Moreover, the constraints are satisfied by the scalar 
fields (\ref{eq:coin1})-(\ref{eq:coin3}) which depend 
on the moduli fields $\phi(x^\mu)$ and $U(x^\mu)$ 
\begin{align}
\label{d1} 
H_1H_1^{\dagger}+H_2H_2^{\dagger} & = c_1\mathbf{1}_N\,, \\
\label{d2} 
\abs{H_3}^2-\Tr(H_2H_2^{\dagger}) & = c_2\,.
\end{align} 

The equation of motion for $W_\mu$ gives 
\begin{equation}
W_{\mu} = -\frac{i}{2c_1}\Bigl[H_{a}\hat D_{\mu}H_a^{\dagger}
-\hat D_{\mu}H_aH_a^{\dagger}\bigr]
+\frac{1}{c_1} A_{\mu}H_2H_2^{\dagger}
\equiv \hat W_{\mu}+\frac{1}{c_1}A_{\mu}H_2H_2^{\dagger}\,,
\end{equation}
where the sum over the index $a=1,2$ is implied (Einstein 
summation convention). 
Plugging this back into \refer{eff} we obtain
\begin{align}
\oper{L}^{(2)} & = \Tr(\hat D_{\mu}H_a\hat D^{\mu}H_a^{\dagger}) -c_1\Tr(\hat W_{\mu}\hat W^{\mu})
 +i A_{\mu}\Bigl(H_3\partial^{\mu}H_3^{\dagger}-\partial^{\mu}H_3H_3^{\dagger} \nonumber \\ & -
\Tr(H_2\hat D^{\mu}H_2^{\dagger}-\hat D^{\mu}H_2H_2^{\dagger})\Bigr)
 -2A_{\mu}\Tr(\hat W^{\mu}H_2H_2^{\dagger})\nonumber \\ &+A_{\mu}A^{\mu}\Bigl(\abs{H_3}^2+\Tr(H_2H_2^{\dagger})-\frac{1}{c_1}\Tr\bigl[(H_2H_2^{\dagger})^2\bigr]\Bigr)
+\partial_\mu H_3^{\dagger}\partial^{\mu}H_3\,.
\end{align}

Next step is to eliminate the auxiliary fields $A_{\mu}$.
By using the following identities derived from the 
constraint \refer{d1} 
\begin{align}
\Tr\bigl[(H_2H_2^{\dagger})^2\bigr] & = c_1\Tr(H_2H_2^{\dagger})-\Tr(H_2H_2^{\dagger}H_1H_1^{\dagger})\,, \\
-2\Tr(\hat W^{\mu}H_2H_2^{\dagger}) & = i \Tr(H_2\hat D_{\mu}H_2^{\dagger}-\hat D_{\mu}H_2H_2^{\dagger}) \nonumber \\
 & +\frac{i}{c_1}\epsilon_{ab}\Tr\bigl[(H_a\hat D_{\mu}H_a^{\dagger}-\hat D_{\mu}H_aH_a^{\dagger})H_bH_b^{\dagger}\bigr]\,,
\end{align}
and by solving equations of motion for $A_{\mu}$, we obtain 
\begin{equation}
A_{\mu} = -\frac{i}{2}\frac{H_3\partial_{\mu}H_3^{\dagger}-\partial_{\mu}H_3H_3^{\dagger}
+\tfrac{1}{c_1}\epsilon_{ab}\Tr\bigl[(H_a\hat D_{\mu}H_a^{\dagger}-\hat D_{\mu}H_aH_a^{\dagger})H_bH_b^{\dagger}\bigr]}
{\abs{H_3}^2+\tfrac{1}{c_1}\Tr(H_2H_2^{\dagger}H_1H_1^{\dagger})}\,.
\end{equation} 
By using the following identities 
\begin{equation}
\Tr\Bigl[\hat D_{\mu}H_a\hat D^{\mu}H_a^{\dagger} -c_1\hat W_{\mu}\hat W^{\mu}\Bigr]
= \frac{1}{2c_1} 
\Tr\Bigl[\oper{D}_{\mu}\oper{H}_{ab}\oper{D}^{\mu}\oper{H}_{ab}^{\dagger}\Bigr]\,,
\end{equation}
where $\oper{H}_{ab} := H_a^{\dagger}H_b$ and ${\mathcal D}_\mu \oper{H}_{ab}=\partial_\mu \oper{H}_{ab}+i[V_\mu, \oper{H}_{ab}]$, and 
\begin{equation}
\epsilon_{ab}\Tr\bigl[(H_a\hat D_{\mu}H_a^{\dagger}-\hat D_{\mu}H_aH_a^{\dagger})H_bH_b^{\dagger}\bigr]
= \Tr\bigl[\oper{H}_{12}^{\dagger}\oper{D}_{\mu}\oper{H}_{12}-\oper{D}_{\mu}(\oper{H}_{12}^{\dagger})\oper{H}_{12}\bigr]\,,
\end{equation}
we obtain a simpler expression for 
$A_{\mu}$ 
\begin{equation}\label{amu}
A_{\mu} = -\frac{i}{2}\frac{H_3\partial_{\mu}H_3^{\dagger}-\partial_{\mu}H_3H_3^{\dagger}
+\tfrac{1}{c_1}\Tr\bigl[\oper{H}_{12}^{\dagger}\oper{D}_{\mu}\oper{H}_{12}-\oper{D}_{\mu}(\oper{H}_{12}^{\dagger})\oper{H}_{12}\bigr]}
{\abs{H_3}^2+\tfrac{1}{c_1}\Tr\bigl[\oper{H}_{12}^{\dagger}\oper{H}_{12}\bigr]}\,.
\end{equation} 
Using (\ref{amu}),  
we can rewrite the effective Lagrangian as an integral 
of a nonlinear sigma model over $y$ 
\begin{multline}\label{eff2}
 \oper{L}_{\rm eff}  = \lineint y\, \biggl[\frac{1}{2c_1}\Tr\Bigl[\oper{D}_{\mu}\oper{H}_{ab}\oper{D}^{\mu}\oper{H}_{ab}^{\dagger}\Bigr]
 +\partial_{\mu}H_3\partial^{\mu}H_3^{\dagger} \\ 
-A_{\mu}A^{\mu}\Bigl(\abs{H_3}^2+\tfrac{1}{c_1}\Tr(\oper{H}_{12}^{\dagger}\oper{H}_{12})\Bigr)\biggr]\,.
\end{multline}

\subsection{Effective Lagrangian for $U$ and $y_0$}

Let us now calculate the effective Lagrangian including only 
fluctuations $U$ and $y_0$ around the coincident 
domain wall solutions \refer{sol1}-\refer{sol3} 
with \refer{coin1} and \refer{coin2}.
The composite fields $\oper{H}_{ab}$ are given as 
($e_0 \equiv e^{m(y-y_0)}$) 
\begin{align}
\oper{H}_{11} & = c_1\frac{e_0e^{-\eta}}{1+e_0e^{-\eta}}\mathbf{1}_N\,, \\
\oper{H}_{12} & = c_1\frac{e_0^{1/2}e^{-\eta/2}}{1+e_0e^{-\eta}}U^{\dagger}\,, \\
\oper{H}_{22} & = c_1\frac{1}{1+e_0e^{-\eta}}\mathbf{1}_N\,.
\end{align}
After some algebra, 
covariant derivatives of these can be rewritten as 
\begin{align}
\oper{D}_{\mu}\oper{H}_{11} & = -\frac{2c_2}{N}e^{-\eta}\sigma \partial_{\mu}y_0 \mathbf{1}_N\,, \\
\oper{D}_{\mu}\oper{H}_{12} & = \frac{2c_2}{N}e^{-\eta}\sigma \sinh\bigl(m(y-y_0)-\eta/2\bigr)\partial_{\mu}y_0 \,U^{\dagger}\nonumber \\
& +\frac{c_1}{2}\frac{\oper{D}_{\mu}U^{\dagger}}{\cosh\bigr(m(y-y_0)-\eta/2\bigl)}\,, \\
\oper{D}_{\mu}\oper{H}_{22} & = \frac{2c_2}{N}e^{-\eta}\sigma \partial_{\mu}y_0 \mathbf{1}_N\,,
\end{align}
where 
\begin{equation}
\frac{\sigma}{m} = \Bigl(1+\frac{c_2}{Nc_1}(1+e_0e^{-\eta})^2/e_0\Bigr)^{-1}\,.
\end{equation}
Putting this into Eq. \refer{amu} we obtain:
\begin{equation}
A_{\mu} = -\frac{i\sigma}{2Nm}\Tr[U\oper{D}_{\mu}U^{\dagger}-U^{\dagger}\oper{D}_{\mu}U]\,.
\end{equation}
Substituting this result into \refer{eff2}, we obtain
\begin{align}
\oper{L}_{\rm{eff}} & = \frac{c_2}{Nm}\lineint y\, \frac{\sigma}{1-\sigma/m}e^{-\eta}\Tr\bigl[\oper{D}_{\mu}U^{\dagger}\oper{D}^{\mu}U\bigr]
+mc_2\lineint y\,e^{-\eta}\sigma\partial_{\mu}y_0\partial^{\mu}y_0
\nonumber \\ & +\frac{c_2}{N^2m^2}\lineint y\, 
\frac{\sigma^2}{1-\sigma/m}e^{-\eta}
\Tr\bigl[U\oper{D}_{\mu}U^{\dagger}\bigr]\Tr\bigl[U\oper{D}^{\mu}U^{\dagger}\bigr]\,.
\end{align}
Finally integration over $y$ is most easily done by 
substituting $x = e^{-\eta}$ and using the identity:
\begin{equation}
\frac{\sigma}{m} = \frac{(x-1)\Bigl(1-\tfrac{c_2}{Nc_1}(x-1)\Bigr)}{x+(x-1)\Bigl(1-\tfrac{c_2}{Nc_1}(x-1)\Bigr)}\,,
\end{equation} 
leading to Eq.(\ref{result}) (including the kinetic term 
for the gauge fields).

\subsection{Effective Lagrangian for all terms: $\phi$, $U$ and $y_0$}

Next we derive the effective Lagrangian 
including all fluctuations.
Plugging the solution \refer{Higgs-g1}-\refer{Higgs-g3} into the Lagrangian \refer{eff2}, we obtain
\begin{align}
\oper{L}_{\rm eff} & =  \lineint y\, \Bigl[\oper{T}_{U}+\oper{T}_{mix}+\oper{T}_{\phi}\Bigr]\nonumber \\
& +\frac{c_1^2}{16c_2}\lineint y\,e^{\eta}\Bigl(1-\frac{\sigma}{m}\Bigr)\Tr\Bigl[\frac{1}{\cosh^2(\hat y)}\oper{D}_{\mu}U^{\dagger} U\Bigr]
\Tr\Bigl[\frac{1}{\cosh^2(\hat y)}\oper{D}^{\mu}U^{\dagger}U\Bigr]\nonumber \\
& -\frac{c_1^2}{16c_2}\lineint y\,e^{\eta}\Bigl(1-\frac{\sigma}{m}\Bigr)\Tr\Bigl[\frac{1}{\cosh^2(\hat y)}\oper{D}_{\mu}\phi\Bigr]
\Tr\Bigl[\frac{1}{\cosh^2(\hat y)}\oper{D}^{\mu}\phi\Bigr]\,,
\end{align}
where
\begin{align}
\oper{T}_{U} & = \frac{c_1}{4} \Tr\biggl\{\oper{D}_{\mu}U^{\dagger}\oper{D}^{\mu}U\frac{1}{\cosh^2(\hat y)}\nonumber \\
& +\oper{D}_{\mu}U^{\dagger}U\sum_{n=1}^{\infty}\frac{(-1)^n}{n!}\oper{L}_{\phi}^n(U^{\dagger}\oper{D}^{\mu}U)
\biggl(\frac{e^{\hat y}}{\cosh(\hat y)}\biggr)^{(n)}
\frac{e^{-\hat y}}{\cosh(\hat y)}\biggr\}\,,
\end{align}
\begin{align}
\oper{T}_{mix} & = - \frac{c_1}{2}\sum_{n=2}^{\infty}\frac{(-1)^n}{n!}\Tr\biggl\{U^{\dagger}\oper{D}_{\mu}U\oper{L}_{\phi}^{n-1}
\Bigl(\oper{D}^{\mu}\phi\Bigr)\Bigl[\Bigl(\frac{1}{\cosh(\hat y)}\Bigr)^{(n)}\frac{1}{\cosh(\hat y)}\nonumber \\ &
+\Bigl(\tanh(\hat y)\Bigr)^{(n)}\tanh(\hat y)
\Bigr]\biggr\}\,,
\end{align}
\begin{align}
\oper{T}_{\phi} & = \frac{c_2}{4}\frac{\sigma/m-2}{1-\sigma/m}e^{-\eta}\partial_{\mu}\eta\partial^{\mu}\eta+\frac{c_1}{4}\Tr\Bigl[
\frac{1}{\cosh^2(\hat y)}\oper{D}_{\mu}\phi\oper{D}^{\mu}\phi\Bigr]
\nonumber \\ &- \frac{c_1}{2}\sum_{n=3}^{\infty}\frac{1}{n!}\Tr\biggl\{\oper{L}_{\phi}^{n-2}
\Bigl(\oper{D}_{\mu}\phi\Bigr)\oper{D}^{\mu}\phi\Bigl[\Bigl(\frac{1}{\cosh(\hat y)}\Bigr)^{(n)}\frac{1}{\cosh(\hat y)}\nonumber \\
& +\Bigl(\tanh(\hat y)\Bigr)^{(n)}\tanh(\hat y)
\Bigr]\biggr\}\,,
\end{align}
and 
\begin{equation}
\hat y = (my-\eta/2)\mathbf{1}_N-\phi\,.
\end{equation}
The only work which remains to be done is to perform the 
integration. 
This, however, turns out to be difficult. 
The main source of difficulty lies in the fact the we 
cannot solve \refer{eq:geneta} explicitly. We 
therefore use some approximation technique, such as 
the Taylor expansion. 
We find it convenient to use as an expansion 
parameter $c_1/c_2$  which determines the width of 
the coincident wall. 
In the lowest order of approximation we see that 
Eq.\refer{eq:geneta} reduces to $e^{-\eta} = 1$ and 
we obtain the effective Lagrangian in Eq.\refer{eq:efflagr}.

\subsection{Complexity of the Effective Lagrangian 
}

The formula \refer{eq:efflagr} illustrates the 
complexity of the interactions between moduli fields 
$\phi$ and $U$ in the general case.
Let us offer some explanation for this complexity. 
The structure, which is new in \refer{eq:efflagr} compared 
to \refer{eq:efflagr0} is of the general form
\begin{equation}
F(\partial_x)\Tr[e^{x\phi}M(y)]\Tr[e^{-x\phi}N(y)]\Big|_{x=0}\,,
\end{equation}
where $M(y)$ and $N(y)$ are some matrix-valued functions, containing either derivatives of $\phi$ or derivatives of $U$.
If we assume that $\phi = mP^{-1}\mathrm{diag}(y_1,\ldots ,y_N)P$ is diagonalizable we can rewrite the above as 
\begin{equation}
\sum_{i,j=1}^NF(m(y_i-y_j))(PM(y)P^{-1})_{ii}(PN(y)P^{-1})_{jj}\,.
\end{equation}
This form suggests that interaction (here represented by function $F$) depends on the relative size of fluctuation of each pairs of walls.
Indeed, notice that if two walls have the same position  $y_i = y_j$ $i\not = j$ (meaning that the expectation values of the fluctuations is the same), the above form reduces to
\begin{equation}
F(0)\Tr(M)\Tr(N)\,,
\end{equation}
which is in a sense trivial, since we already encountered this kind of terms in \refer{result}.
Thus, the new kind of complexity in our result \refer{eq:efflagr}
can be understood as a manifestation of the fact, that the interaction does not depend only on various moments of the fluctuation as in 
\refer{eq:efflagr0} but also on their relative size.

\section{Geometry of the three-flavor model}
\label{ap:3flavor}

Quantum numbers of three-flavor model are shown in 
Tab.~\ref{table:03}.
\begin{table}
\begin{center}
\begin{minipage}[b]{0.49\linewidth}
\centering
\begin{tabular}{c|cc|c}
\hline
 & $U(1)_1$ & $U(1)_2$ & mass\\ \hline
$H_{1}$ & 1 & 0 & $m$\\ 
$H_{2}$ & 1 & $1$ & $0$\\ 
$H_{3}$ &  0 & $-1$ & $0$\\ 
$\sigma_1$ & 0 & 0 & 0\\
$\sigma_2$ & 0 & 0 & 0 \\
\hline
\end{tabular}
\end{minipage}
\end{center}
\caption{Quantum numbers of the $U(1)_1\times U(1)_2$ 
three-flavor model.}
\label{table:03}
\end{table}
Since we are interested in the domain wall solutions, 
it is enough to consider the strong gauge coupling limit 
where gauge theories become non-linear sigma models 
(NLSM), 
whose target space is defined as an intersection of two 
spaces as 
\begin{eqnarray}
\left(\mathbb{C} \times \mathbb{C}P^1\right) \cap 
\left(\mathbb{C} \times \mathbb{H}^2\right) 
\simeq \mathbb{C}P^1.
\end{eqnarray}
Here $\mathbb{H}^2$ stands for the two dimensional 
hyperbolic plane. 
In the above expression, The $\mathbb{C}P^1$ and 
$\mathbb{H}^2$ are defined by 
\begin{eqnarray}
\mathbb{C}P^1 = \left\{ (H_1,H_2)\ \big|\ 
|H_1|^2 + |H_2|^2 = c_1,\ H_{1,2}\in 
\mathbb{C}\right\}/U(1)_1,\\
\mathbb{H}^2 = \left\{ (H_2,H_3)\ \big|\ 
|H_2|^2 - |H_3|^2 = -c_2,\ H_{2,3}\in 
\mathbb{C}\right\}/U(1)_2.
\end{eqnarray}
This space is isomorphic to $\mathbb{C}P^1$ but it is a 
squashed sphere. 
The metric can be read from the K\"ahler potential 
\begin{eqnarray}
K = e^{-V_1}|H_1|^2 + e^{-V_1-V_2}|H_2|^2 
+ e^{V_2}|H_3|^2 + c_1 V_1 - c_2 V_2.
\end{eqnarray}
One can eliminate the real superfields $V_1$ and $V_2$, and find the following expression
of the K\"ahler potential with respect to the gauge invariant inhomogeneous coordinate $\varphi$ as
\begin{eqnarray}
K &=& c_1 \left[ f + \log\left(|\varphi|^{-2}+f^{-1}\right) - \lambda \log f\right],\\
f &=& \frac{1}{2}\left(\lambda - |\varphi|^2  + \sqrt{4|\varphi|^2(1+\lambda) + \left(\lambda - |\varphi|^2\right)^2}\right),\\
\varphi &=& \frac{H_2H_3}{c_1H_1},\quad
\lambda \equiv \frac{c_2}{c_1}.
\end{eqnarray}
Note that this manifold is singular at $\varphi=0$ when $c_2=0$.
This can be understood in two different ways. The first one is to realize the fact that
$U(1)_2$ gauge symmetry is restored at that point. This is because
$|H_2| = |H_3| = 0$ holds there. Namely, the additional massless degrees of freedom
should be taken into account. The second way is more straightforward.
Let us calculate the scalar curvature in the 
vicinity of $\varphi = 0$.
To this end, first we change the coordinate by $\varphi = e^{i\Phi}\tan\frac{\Theta}{2}$
($0 \le \Theta \le \pi,\ 0 \le \Phi \le 2\pi$), then we 
obtain
\begin{eqnarray}
R = \frac{8 (8+9 \lambda )}{9} \frac{1}{c_2} + {\mathcal O}(\Theta^2).
\end{eqnarray}
From this it is clear that the scalar curvature at $\varphi = 0$ becomes infinity when $c_2 = 0$.

 \end{document}